\documentclass[10pt,final,doublecolumn]{IEEEtran}
\IEEEoverridecommandlockouts
\usepackage{multirow}
\usepackage{amsmath}
\usepackage{amssymb}
\usepackage{amsthm}
\usepackage{mathrsfs}
\usepackage{latexsym}
\usepackage{graphicx}
\usepackage{bbding}
\usepackage{indentfirst}
\usepackage{cases}
\usepackage{supertabular}
\usepackage{algorithm,algorithmic}
\usepackage{subeqnarray}
\usepackage{color}
\usepackage{bm}
\usepackage{stfloats}
\usepackage{subfigure}
\usepackage{array}
\usepackage{algorithm,float}
\usepackage{booktabs}

\IEEEoverridecommandlockouts
\allowdisplaybreaks[4]



\begin{document}
\title{Modeling and Analysis for Multiple-Layer LEO Satellite Internet of Things Constellations}
\author{\IEEEauthorblockN{
Ming Ying, Xiaoming Chen, Qiao Qi, and Yichao Xu}

\thanks{Ming Ying, Xiaoming Chen, and Yichao Xu are with the College of Information Science and Electronic Engineering, Zhejiang University, Hangzhou 310027, China (e-mail:\{ming\_ying, chen\_xiaoming, yichao\_xu\}@zju.edu.cn). Qiao Qi is with the School of Information Science
and Technology, Hangzhou Normal University, Hangzhou 311121, China (email: qiqiao@hznu.edu.cn).}
}\maketitle

\begin{abstract}
    To provide multiple-satellite coverage for global Internet of Things (IoT), a low Earth orbit (LEO) satellite IoT constellation usually contains multiple-layer orbits with different altitudes. However, the performance of multiple-layer LEO satellite IoT constellations under practical Rician fading satellite channels remains unknown due to complex theoretical modeling and intractable mathematical analysis. To address these challenges, this paper proposes a stochastic geometry-based modeling and analysis framework for multiple-layer LEO satellite IoT constellations, integrating Rician channel modeling and Cox point processes. Specifically, we introduce a novel channel approximation method to overcome the intractable expressions caused by the Rician fading. Building on this method, we derive exact closed-form expressions for key performance metrics, including connectivity probability, coverage probability, and transmission rate, especially in the case of IoT short-packet transmission. Extensive simulation results validate the accuracy and effectiveness of the proposed model and reveal significant design insights. The results not only provide new theoretical perspectives for modeling and analysis of LEO satellite IoT constellations but also offer practical guidance for system deployment and optimization. 
\end{abstract}

\providecommand{\keywords}[1]{\textbf{\textit{Index Terms---}} #1}

\section{Introduction}
     Internet of Things (IoT) has witnessed exponential growth, enabling ubiquitous connectivity across diverse applications, such as intelligent transportation, precision agriculture, and environmental monitoring \cite{iot1}-\cite{iot3}. These applications increasingly demand ubiquitous, reliable, and low-latency connectivity, especially in scenarios involving mobile or geographically isolated devices. However, terrestrial networks often struggle to meet these requirements due to limitations in infrastructure deployment and geographical reach. In particular, vast regions such as oceans, and deserts remain underserved by ground-based systems, where deploying cellular IoT infrastructure is neither economically viable nor technically feasible \cite{siot1}. To address this gap, low Earth orbit (LEO) satellite networks have emerged as a promising complementary solution, offering global coverage, reduced latency, and robust connectivity for distributed IoT devices \cite{siot2}. Unlike geostationary systems, LEO satellites operate at lower altitudes, which significantly cuts propagation delay and power consumption, offering key advantages for energy-constrained IoT deployments \cite{siot3}. Nevertheless, the dynamic characteristic of LEO constellations introduces challenges such as intermittent visibility and complex handover management, necessitating carefully designed multi-layer architectures to ensure seamless service continuity.

	 Given the substantial investment and strategic importance of LEO satellite IoT constellations, accurate performance modeling and prediction become prerequisites for viable system design and commercialization. A rigorous theoretical framework is therefore essential to guide constellation design and optimize resource allocation \cite{dsn}. {In this context, stochastic geometry has been proven to be a powerful tool for modeling large-scale wireless networks, providing tractable expressions for key performance indicators such as coverage, capacity, and reliability.} Its ability to capture spatial randomness makes it particularly suitable for analyzing LEO satellite IoT constellations.

     In recent years, there have been growing researches on applying stochastic geometry to analyze the performance of LEO satellite networks, with the Poisson point process (PPP) emerging as a predominant modeling approach in LEO satellite IoT \cite{dsga}-\cite{mlcc}. Building on this foundation, \cite{dsga} introduced an equivalent two-dimensional PPP model incorporating Nakagami-m fading channels to streamline performance analysis for LEO satellite constellations. {Subsequent studies extended this framework to evaluate critical performance metrics. In \cite{pmc}, the authors derived outage probability and ergodic capacity for uplink transmission by utilizing PPP models with standard antenna patterns. Moreover, the authors in \cite{bpp1} characterized user fairness and reliability via the distribution and moments of conditional coverage probability under a homogeneous PPP. For mega LEO satellite constellation design, the authors in \cite{bpp3} proposed a theoretical PPP-based framework for capacity analysis and presented an optimal constellation design algorithm minimizing the satellite constellation costs while ensuring global coverage. Complementing these efforts, the authors in \cite{bpp2} developed an analytical PPP framework specifically for LEO satellite network backhaul performance evaluation.} To better capture the real-world spatial distribution of LEO satellites across varying latitudes, \cite{mlcc} employed a non-homogeneous PPP with an appropriately defined intensity function. Subsequently, the researchers have advanced binomial point process (BPP) modeling for multiple-layer LEO satellite constellations, with \cite{bpp} pioneering closed-form coverage and rate expressions while introducing the concept of effective satellite counts to connect theoretical models with practical design considerations. This foundation has spurred numerous BPP-based investigations \cite{sna}-\cite{asga4}. In \cite{sna}, the authors validated the accuracy of BPP modeling for LEO satellite constellations by establishing the statistical equivalence between BPP-modeled LEO constellations and Fibonacci lattice distributions. Moreover, \cite{asga} proposed a stochastic geometry-based framework modeling the LEO satellite constellation as a BPP model on a spherical surface. By approximating satellite distances and incorporating shadowed-Rician (SR) fading channels, the closed-form expressions for transmission success probability were derived. {Similarly, \cite{asga2} established an exact BPP-based framework for LEO satellite downlink systems and derived outage probability and throughput maximization solutions through the innovative Poisson approximation and iterative optimization.} Since oversimplified assumptions like infinite terrestrial areas and omnidirectional coverage often employed in LEO satellite IoT, a BPP-based framework incorporating finite-area effects, Earth curvature, and dual-link Nakagami-m/SR fading was proposed in \cite{asga3} to enable realistic performance evaluation under operational constraints. Moreover, a BPP framework for multi-altitude LEO satellite constellations was developed in \cite{asga4} and revealed diminishing returns from constellation overprovisioning. Beyond connectivity, the BPP framework has also been extended to specialized domains such as reliability, security, and heterogeneous communications. For instance, \cite{A2} quantified multi-hop connection reliability in hybrid satellite-terrestrial networks under different relay strategies. In security analysis, \cite{A3} employed a BPP to model node locations and study covert communication under adversarial detection, while \cite{A4} derived tractable physical-layer security metrics using a BPP-based constellation model. The versatility of BPP is further demonstrated in heterogeneous systems, where \cite{A5} applied it to jointly optimize coverage and rate in terahertz/RF multi-band satellite networks. 

     Although both PPP and BPP models have advanced LEO satellite network analysis, they fundamentally overlook critical geometric constraints, particularly the predetermined orbital paths governing each satellite's operation. Fortunately, recent works in \cite{cox1}-\cite{cox3} address this issue through a satellite Cox point process (CPP) that jointly models satellites and their exclusive orbits via Cox structures. However, current researches still exhibit several critical limitations. Firstly, the analysis of \cite{cox1}-\cite{cox3} is mainly based on Nakagami-m channel with $m=1$, i.e., Rayleigh fading model, which fails to capture the dominant line-of-sight (LoS) components inherent in LEO satellite communication channels. The Rician fading model, which properly accounts for both LoS and scattered components, introduces significant mathematical complexity that remains largely unaddressed in the existing literature. This gap fundamentally limits the accuracy of performance analysis for LEO satellite IoT constellations operating under realistic channel conditions. Secondly, while existing researches mainly focus on single-layer LEO satellite IoT constellation, practical operational networks increasingly adopt multiple-layer architectures to achieve global coverage and enhanced capacity. This critical gap between theoretical models and real-world deployments leads to significant inaccuracies in performance evaluation, particularly in characterizing inter-layer interference and handover dynamics. Thirdly, current approaches predominantly address satellite constellation properties while overlooking inherent characteristics of IoT devices. Specifically, IoT devices typically employ short-packet communications, fundamentally limiting their achievable transmission rates below the Shannon capacity bound. Motivated by these research deficiencies, this paper establishes a comprehensive modeling and analysis framework for LEO satellite IoT constellations.
     
     {Beyond spatial modeling considerations, significant efforts have also been made to incorporate more realistic channel models into the stochastic geometry analysis of satellite networks. Notably, several recent studies have employed SR fading model to better capture the signal fluctuations in realistic LEO satellite channels. Specifically, \cite{asga2} derived outage probability and optimized throughput for LEO satellite systems under SR fading using stochastic geometry. Moreover, \cite{D2} developed an analytical framework for multi-beam LEO satellite downlink transmission, simplifying SINR analysis under correlated SR channels. For non-orthogonal multiple access (NOMA) in satellite networks, \cite{D3} analyzed outage behaviors and revealed the critical impact of imperfect interference cancellation over SR links. Furthermore, \cite{D4} applied the large deviations principle to obtain concise outage probability bounds for SR-faded satellite-terrestrial channels. These contributions are crucial for advancing the field and demonstrate the importance of accurate channel characterization. Motivated by these works but focusing on the specific deployment scenarios considered herein, we note that while the SR model offers a more general characterization for scenarios with LoS fluctuation, the Rician model provides a tractable and accurate approximation for IoT devices deployed in open environments (e.g., farmland, deserts, oceans) where a stable LoS path is predominant. Therefore, this work adopts the Rician channel model to maintain analytical tractability while capturing the essential propagation characteristics relevant to our target application scenarios. }
     
     Specifically, the contributions are listed as below:
     
     \begin{enumerate}
        \item We provide a novel channel approximation method for Rician channel modeling by using finite General Dirichlet series, enabling tractable performance analysis while maintaining high accuracy. Simulation results demonstrate that our method achieves excellent matching with the actual Rician channel characteristics, providing a valuable tool for channel modeling of LEO satellite IoT constellations.
        \item We develop an analytical framework to evaluate the performance of multi-layer LEO satellite constellations in IoT scenarios, where each device can potentially access any satellite across different orbital layers. Specifically, we derive key performance metrics, including connectivity probability, coverage probability and transmission rate, by utilizing a geometry-based model that incorporates the Laplace transform of the aggregate interference power.            
        \item We conduct extensive simulations to verify the accuracy of the derived expressions and quantitatively analyze the impacts of key parameters on the performance of the LEO satellite IoT constellation, including satellite densities, orbit densities, and the orbit altitudes. Furthermore, we demonstrate the geometric accuracy of the proposed model by comparing its predictions with realistic constellation deployments, showing strong agreement between theoretical and practical scenarios.     
    \end{enumerate}

   The remainder of this paper is organized as follows. In Section II, we introduce the system model of LEO satellite IoT constellations, including the orbital elements, channel model and transmission model. In Section III, we first derive the distance distribution between the LEO satellite and IoT devices under CPP model. Based on this, we present the expressions of connectivity probability, coverage probability and transmission rate of LEO satellite IoT constellations. Next, extensive simulation results are provided in Section IV to verify the accuracy of the derived expressions and investigate the effects of key parameters on the performance of the LEO satellite IoT constellations. Finally, Section V concludes the paper.
   
   {Specifically, the notations of this paper are listed in Table \ref{not}.
   
   \begin{table}[h]
   	\label{not}
   	\scriptsize
   	\centering
   	\caption{Notations of This Paper}
   	\begin{tabular}{|c|c|}
   		\hline
   		Symbol & Description \\\hline\hline
   		$\mathbb{E}[X]$ & Expectation of random variable $X$ \\\hline
   		$\Pr(X > x)$ & Complementary cumulative distribution function of $X$ \\\hline
   		$\bigcup$ & Union operator \\\hline
   		$Q_1(a,b)$ & First-order Marcum $Q$-function with parameters $a$ and $b$ \\\hline
   		$I_0(\cdot)$ & Modified Bessel function of the first kind \\\hline
   	\end{tabular}
   \end{table}}


\section{System Model}
     
     \begin{figure}
        \centering
        \includegraphics [width=0.5\textwidth] {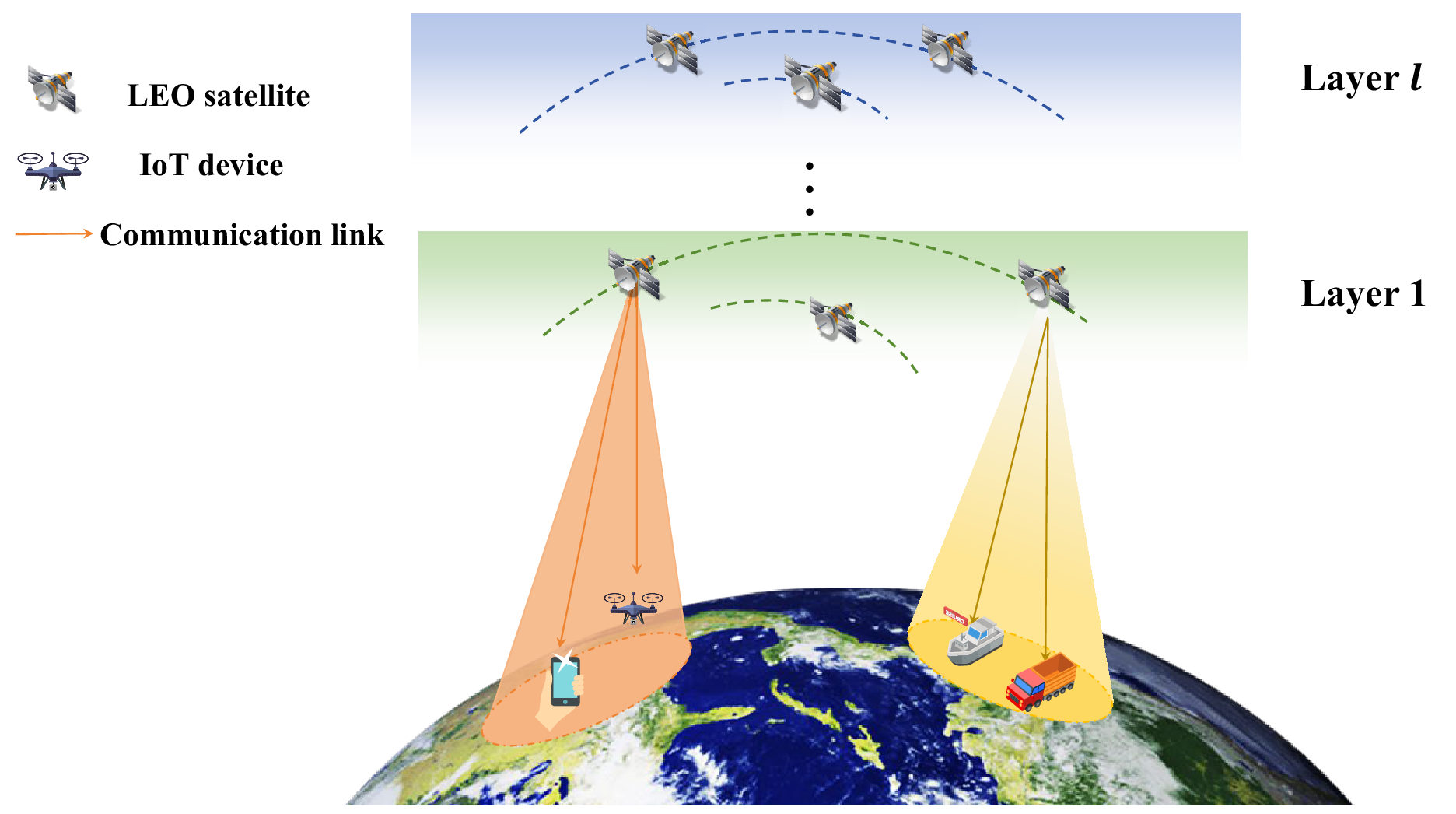}
        \caption {System model of multiple-layer LEO satellite IoT constellation.}
        \label{sys}
    \end{figure}

    {This paper considers a multiple-layer LEO satellite IoT constellation where a massive number of IoT devices are deployed within open or rural environments \cite{siot}, typically characterized by a dominant and stable line-of-sight (LoS) path to the satellites, as illustrated in Fig. 1.} To provide multiple-satellite coverage, LEO satellites are deployed in $L$ layers with different orbit altitudes. Each IoT device communicates with one LEO satellite which provides the maximum received signal power. In what follows, we introduce orbital elements, geometric model, and channel model, for the multiple-layer LEO satellite IoT constellation\footnote{The large carrier frequency offset (CFO) caused by high mobility of the LEO satellite is compensated in advance according to the deterministic trajectory of the LEO satellite \cite{cfo1}, \cite{cfo2}.}.
    \subsection{Orbital Elements}
    \begin{figure}
        \centering
        \includegraphics [width=0.5\textwidth] {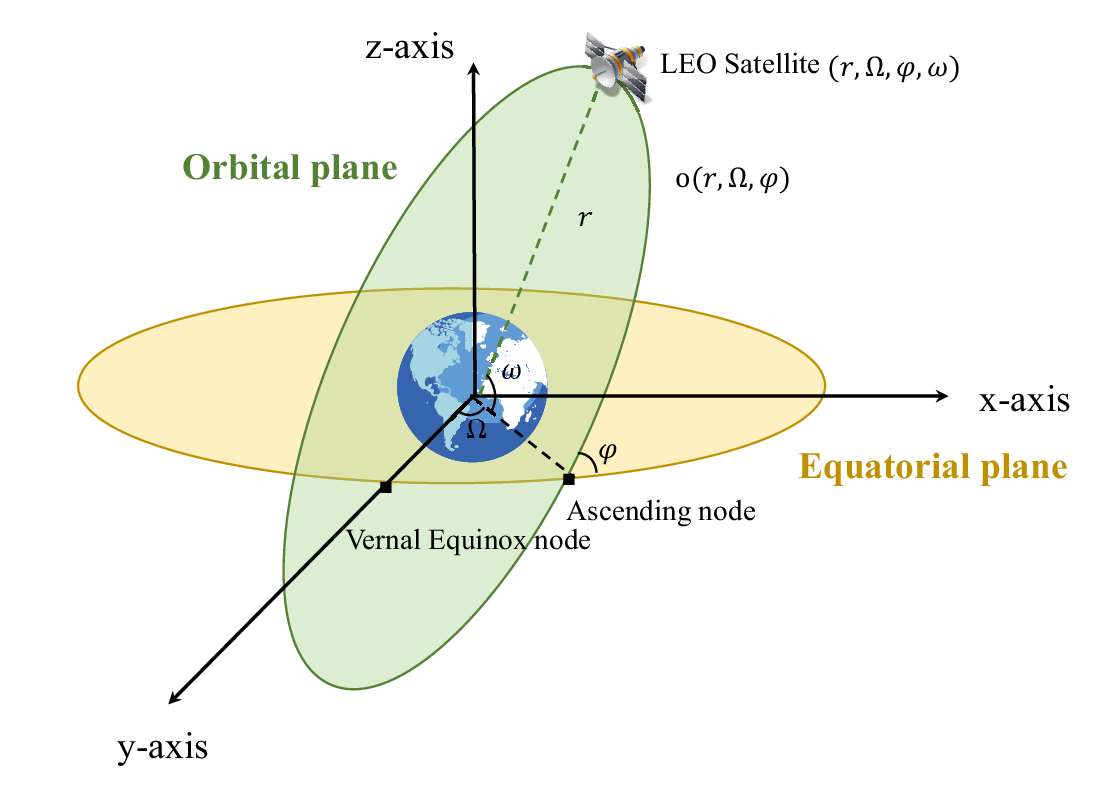}
        \caption {Illustration of orbital elements of LEO satellite IoT constellation.}
        \label{satcons}
    \end{figure}
    To precisely characterize the positions of LEO satellites in space, we employ the six classical orbital elements within the Earth-Centered Inertial (ECI) reference frame. Specifically, the six orbital elements are defined as follows:
    \begin{enumerate}
        \item Semi-major $a_s$: Half the length of the orbit’s major axis. 
        \item Eccentricity $e$: The eccentricity of the orbital ellipse.
        \item Inclination angle $\varphi$: The angle measured counterclockwise from the equatorial plane to the satellite orbital plane at the ascending node. An inclination angle of more than 90$^\circ$ indicates a retrograde orbit. 
        \item Right ascension of ascending node $\Omega$: Angle measured counterclockwise from the vernal equinox node to the ascending node in the equatorial plane. 
        \item Argument of Periapsis $\omega_p$: Angle measured counterclockwise from the ascending node to the perigee, i.e., the point when the satellite operates closest to Earth, measured along the orbital plane.
        \item True anomaly $\upsilon$: Angle measured counterclockwise from the perigee to the satellite position.
    \end{enumerate}
    With these six orbit elements, the position of each satellite can be accurately described. For analytical tractability, we adopt the commonly used circular orbit assumption \cite{ass1}-\cite{ass3}, as visualized in Fig. \ref{satcons}. Under this circular orbit assumption, the semi-major $a_s$ of the LEO satellites in layer $l$ is equal to the orbit radius, i.e., $a_s = r_l$, where $r_l$ denotes the orbit radius of layer $l$. Moreover, the eccentricity of the orbit is $e=0$. Under such an assumption, the orbit in layer $l$ can be uniquely represented by three parameters $o(r_l,\Omega,\varphi)$. Furthermore, the angular position of a satellite relative to the ascending node $\omega$ is determined by $\omega=\omega_p+v$. Based on these relationships, the Cartesian coordinates of a LEO satellite in layer $l$ can be expressed as \cite{position}
    \begin{equation}\label{cor}
    \left(
    \begin{aligned}
        x_{\mathrm{sat}} & \\
        y_{\mathrm{sat}} & \\
        z_{\mathrm{sat}} & \\
    \end{aligned}\right) = r_l\left(
    \begin{aligned}
        \mathrm{cos}\omega\mathrm{cos}\Omega&-\mathrm{sin}\omega\mathrm{sin}\Omega\mathrm{cos}\varphi  \\ \mathrm{cos}\omega\mathrm{sin}\Omega&+\mathrm{sin}\omega\mathrm{cos}\Omega\mathrm{cos}\varphi   \\
        &\mathrm{sin}\omega\mathrm{sin}\varphi    \\
    \end{aligned}\right).
    \end{equation} 
    Consequently, the spatial position of LEO satellite $m$ in layer $l$ can be uniquely characterized by the quadruple $(r_l, \Omega_m, \varphi_m, \omega_m)$, where $\Omega_m, \varphi_m,$ and $\omega_m$ represent the right ascension of ascending node of satellite $m$, inclination angle of $m$, and the angular position of satellite $m$ relative to the ascending node, respectively.
\subsection{Geometric Model}
    Then, we study the geometric model for the distribution of LEO satellites. According to the characteristics of multiple-layer LEO satellite constellations \cite{cox1}-\cite{cox3}, we use CPP to model the LEO satellite IoT constellation. Based on this, we firstly characterize the orbital distribution of layer $l$ by utilizing a rectangular parameter space defined as $\mathcal{C}_l = r_l \times [0,\pi) \times [0,\pi)$.
    Thus, the satellite orbits are modeled as a PPP $\mathcal{Z}_l$ with an intensity function given by $\frac{\lambda_l\sin\varphi}{2\pi}$, where $\lambda_l$ denotes the average number of orbits in layer $l$.
    
    For each point $Z_{l,i}=(r_l, \Omega_i, \varphi_i)\in\mathcal{Z}_l$, it can be mapped to an orbit $o(r_l,\Omega_i, \varphi_i)$ in the Euclidean space $\mathbb{R}^3$. Then, an orbit process $\mathcal{O}_l$ of layer $l$ can be constructed based on $\mathcal{Z}_l$ as
    \begin{equation}\label{gml}
            \mathcal{O}_l= \bigcup_{Z_{l,i}\in\mathcal{Z}_l}o(r_l, \Omega_i, \varphi_i).
     \end{equation}

    Building upon the orbital model, we characterize the satellite distribution within each orbit. Conditioned on the orbit process $\mathcal{O}_l$, the spatial distribution of satellites along a given orbit $o(r_l, \Omega_i, \varphi_i)$ follows a homogeneous PPP (HPPP) $\psi_i$ with intensity $\frac{\mu_l}{2\pi r_l}$, where $\mu_l$ represents the satellite density in layer $l$. Under the assumption of independent orbital distributions across layers, the process of the whole LEO satellite IoT constellation can be represented as 
    \begin{equation}\label{Psi}
            \Psi=\sum_{l=1}^L \Psi_l=\sum_{l=1}^L \sum_{Z_{l,i}\in\mathcal{Z}_l}\psi_i.
     \end{equation} 
    where $\Psi_l$ denotes the process for LEO satellite IoT constellation of layer $l$. For the distributions of IoT devices, we assume that the IoT devices are uniformly distributed on the surface of Earth $\{(x_{\mathrm{dev}},y_{\mathrm{dev}},z_{\mathrm{dev}})|x_{\mathrm{dev}}^2+y_{\mathrm{dev}}^2+z_{\mathrm{dev}}^2=r_e^2\}$ \cite{cox1}, where $x_{\mathrm{dev}}$, $y_{\mathrm{dev}}$, and $z_{\mathrm{dev}}$ denote the Cartesian coordinates of the IoT devices, $r_e$ is the radius of Earth. Notice that the locations of IoT devices are independent of those of LEO satellites.

    \subsection{Channel Model}
         {According to the propagation characteristics, the LEO satellite channel usually includes two components, namely LoS and non-line of sight (NLoS) \cite{model1},\cite{model3}. This work specifically adopts the Rician fading model for its applicability to open-area IoT deployment scenarios. In such environments, IoT devices typically experience a dominant and stable LoS component with minimal shadowing effects.} Specifically, the large-scale fading factor $g_{l,m}$ of the satellite channel between the LEO satellite $m$ in layer $l$ and IoT device $X_U$ can be expressed as
        \begin{equation}\label{gml}
            g_{l,m} = \sqrt{(\frac{c}{4\pi f d_{l,m}})^2 G_lG_{\mathrm{dev}}\frac{1}{r_0}}=g_t\sqrt{G_ld_{l,m}^{-2}},
        \end{equation}
        where $(\frac{c}{4\pi f d_{l,m}})^2$ denotes the free space loss of the channel with $c$ being the light speed, $f$ being the carrier frequency, 
        and $d_{l,m}$ being the distance between the satellite $m$ of layer $l$ and the device $X_U$. For clarity, $g_t=\sqrt{(\frac{c}{4\pi f} )^2 G_{\mathrm{dev}}\frac{1}{r_0}}$ is adopted to simplify the expressions. Without loss of generality, we consider the device $X_U$ at the north pole as a typical device, then $d_{l,m}$ can be further derived as
        \begin{equation}\label{dlm}
            \begin{aligned}
               d_{l,m} &= \sqrt{r_l^2+r_e^2-2r_lr_e\sin\varphi_m\sin\omega_m}\\
               & = \sqrt{r_l^2+r_e^2-2r_lr_e\cos\overline{\varphi}_m\cos\overline{\omega}_m}=f_{l,\overline{\varphi}_m}(\overline{\omega}_m),
            \end{aligned}     
        \end{equation}
        where $\overline{\varphi}_m=\frac{\pi}{2}-\varphi_m$ and $\overline{\omega}_m=\frac{\pi}{2}-\omega_m$ are denoted as the complementary angles of $\varphi_m$ and $\omega_m$, respectively. $G_{\mathrm{sat},l}$ represents the transmit antenna gain of the LEO satellites in layer $l$, which can be expressed as
    \begin{equation}\label{Gl}
        G_{l} = \left\{
        \begin{aligned}
           &G_{\mathrm{sat},l}, \quad \theta_{l,m}\leq \beta_l\\
           &1, \quad \theta_{l,m}> \beta_l\\
        \end{aligned}\right.,
    \end{equation}
    where $\theta_{l,m}$ is defined as the angle-of-departure (AoD) for the link between the satellite $m$ and the IoT device $X_U$, $\beta_l$ denotes the maximum coverage angle of satellites in layer $l$. The transmit antenna gain $G_{l} = G_{\mathrm{sat},l}$ if the IoT device is in the coverage area of the LEO satellite, otherwise $G_{l} = 1$. $G_{\mathrm{dev}}$ denotes the receive gain of the devices, and $r_0$ denotes the rain attenuation coefficient of the channel whose power gain in dB $r_0^{\mathrm{dB}} = 20\log_{10}r_0$ follows log-normal distribution, i.e., $\ln{(r_0^{\mathrm{dB}})}\sim \mathcal{N}(\mu_r,\sigma_r^2)$. 
    
    For the small scale fading, the probability density function (PDF) and the complementary cumulative distribution function (CCDF) of Rician distributed $|h_{l,m}|^2$ can be expressed respectively as 
    \begin{equation}\label{pdfh}
        \begin{aligned}
            f_{|h_{l,m}|^2}(x) = \frac{1+K}{\Omega_s}&\exp{\Big(-K-\frac{(1+K)x}{\Omega_s}\Big)}\\
            &\times I_0\Big(2\sqrt{\frac{K(1+K)x}{\Omega_s}}\Big),\\
        \end{aligned}
    \end{equation}
    and
    \begin{equation}\label{cdfh}
            \mathcal{F}'_{|h_{l,m}|^2}(x) = Q_1\Big(\sqrt{2K},\sqrt{\frac{2(1+K)x}{\Omega_s}}\Big),
    \end{equation}
    where $K$ denotes the Rician factor which corresponds to the ratio of LoS component and NLoS component, $\Omega_s$ represents the average channel gain calculated as $\Omega_s = \mathbb{E}[|h_{l,m}|^2]$, $Q_1(a,b) = \int_{b}^{\infty}x\exp(\frac{x^2+a^2}{2})I_0(ax) dx $ is
    Marcum-Q function. For clarity, the average channel gain $\Omega_s$ is usually adopted for 1 \cite{omega}. 
    
    Under the maximum received power criterion for satellite selection, the signal to interference plus noise ratio (SINR) $\gamma_{l,m}$ at the typical device $X_U$ when connected to satellite $m$ in layer $l$ is given by
    \begin{equation}\label{gamlm}
            \gamma_{l,m} = \frac{p_lg_{l,m}^2|h_{l,m}|^2}{I_{m}+\sigma_0^2},
    \end{equation}
    where $p_l$ denotes the transmit power of LEO satellites in layer $l$, $\sigma_0^2$ denotes the variance of additive white Gaussian noise (AWGN), $I_{m}$ denotes the total interference from the other satellites. In the following, we propose a modeling and analysis framework to evaluate the performance of the multiple-layer LEO satellite IoT constellation.
    

    \section{Coverage and Rate Analysis}
    In this section, we develop a modeling and analysis framework for evaluating the performance of LEO satellite IoT constellation. {Firstly, we derive the distance distribution between IoT devices and their serving satellites, accounting for multiple-layer orbital geometries.} Then, we establish the connectivity probability considering both spatial distributions and channel conditions. Building on these, we formulate comprehensive coverage probability and transmission rate expressions that capture the essential characteristics of LEO satellite IoT transmission, including interference patterns and fading effects across the constellation.

    Given the satellite selection criterion that IoT devices connect to the LEO satellite offering maximum received signal power, the derivation of coverage probability becomes computationally intractable due to the complex spatial distribution of satellite distances. To overcome this challenge, we employ the law of total probability to decompose the coverage probability into tractable components, yielding
    \begin{equation}\label{covgama1}
        \begin{aligned}
             \mathcal{F}'_{\gamma_\star}(\gamma_{\mathrm{th}})&=\Pr(\gamma_{\star}>\gamma_{\mathrm{th}})\\
             &=\sum_{l=1}^{L}\Pr(\gamma_{\star}>\gamma_{\mathrm{th}}|\mathcal{A}_l)\Pr(\mathcal{A}_l), \\
        \end{aligned}        
    \end{equation}
    where $\gamma_{\star}$ denotes the SINR of the link between the IoT device and the connected LEO satellite, $\gamma_{\mathrm{th}}$ denotes the SINR threshold, and $\mathcal{A}_l$ presents the event that the connected satellite is in layer $l$. Then, the connectivity probability $\Pr(\mathcal{A}_l)$ can be derived as
        \begin{equation}\label{al}
        \begin{aligned}
             \Pr(\mathcal{A}_l)&=\Pr(p_ig_i>p_lg_l,\forall i\in[L]/l)\\
            & =\Pr(D_i>\sqrt{\tau_{i,l}}D_l,\forall i\in[L]/l)\\
            & = \mathbb{E}_{D_l}[\Pr(D_i>\sqrt{\tau_{i,l}}v,\forall i\in[L]/l|D_l=v)],
        \end{aligned}        
    \end{equation}
    where $\tau_{i,l}=p_iG_i/p_lG_l$, $D_l$ denotes the shortest distance between the LEO satellite of layer $l$ and the IoT devices, $[L]$ denotes the set of integers smaller than $L$. Noticing that the process of the constellation in each layer $\Psi_1, \Psi_2, \cdots, \Psi_L$ are independent, we have
     \begin{equation}\label{al2}
             \Pr(\mathcal{A}_l)
             = \mathbb{E}_{D_l}\left[\prod_{\substack{m=1\\ m\neq l}}^{L}\Pr(D_m>\sqrt{\tau_{m,l}}v)\right].
    \end{equation}
    Building on this foundation, we need to derive the distance distribution under specific layer $l$.

    \subsection{Distance Distribution}
        In layer $l$, the connected satellite is the nearest satellite to the IoT devices, i.e., no other LEO satellites in layer $l$ can be closer than $D_l$. Note that when the nearest satellite is in the serving area of layer $l$, we have $r_{l,\min}<D_l<r_{l,\max}$, where $r_{l,\min}=r_l-r_e$, and $r_{l,\max}=r_l\cos\beta_l-\sqrt{r_1^2\cos^2\beta_l-r_l^2+r_e^2}$. From this foundation, we analyze the distribution of distance $\Pr\left(D_l>d\right)$. Specifically, for $d<r_{l,\min}$, $\Pr\left(D_l>d\right)=1$ as the minimum communication distance between the LEO satellite in layer $l$ and the IoT devices is $r_{\min}$. Then, for $r_{l,\max}>d>r_{l,\min}$, according to the properties of CPP \cite{cox1}, the distribution of $D_l$ can be obtained as
        \begin{equation}\label{dis}
            \begin{aligned}
             \Pr\left(D_l>d\right)&\overset{(a)}{=}\Pr{(D_{l,j}>d, \forall X_j\in\Psi_l)}\\
             &\overset{(b)}{=}\Pr{(D_{l,j}>d, \forall X_j\in\psi_i, \forall Z_i\in\mathcal{Z}_l)}\\ &=\Pr{\left(\prod_{Z_i\in\mathcal{Z}_l}\Pr\left(\prod_{X_j\in\psi_i}D_{l,j}>d|\mathcal{Z}_l\right)\right)},
            \end{aligned}
        \end{equation}
        where $(a)$ is derived from the fact that all the distances should be larger than $d$ for all the LEO satellites in layer $l$, and $(b)$ is derived according to the CPP in (\ref{Psi}). As $\psi_i$ follows HPPP, we have 
        \begin{equation}\label{dis2}
            \begin{aligned}
\Pr\left(\prod_{X_j\in\psi_i}D_{l,j}>d|\mathcal{Z}_l\right)=e^{-\frac{\mu_l}{\pi}\arcsin(\sqrt{1-\cos^2{\xi_{l}(d)\csc^2\varphi_i}})}.
            \end{aligned}
        \end{equation}
        where $\xi_l(d)$ is given by
        \begin{equation}\label{xid}
            \xi_l(d) = \arccos(\frac{r_e^2+r_l^2-d^2}{2r_er_l}).
        \end{equation}
        Based on (\ref{dis2}) and probability generating functional (PGFL) of the PPP \cite{pgfl}, (\ref{dis}) can be further calculated as
        \begin{equation}\label{dis3}
            \begin{aligned}
             \Pr&\left(D_l>d\right)
             =\Pr{\left(\prod_{Z_i\in\mathcal{Z}_l}e^{-\frac{\mu_l}{\pi}\arcsin(\sqrt{1-\cos^2{\xi_l(d)\csc^2\varphi_i}})}\right)}\\
             &=e^{-\lambda_l\int_0^{\xi_l(d)}\left(1-e^{-\frac{\mu_l}{\pi}\arcsin(\sqrt{1-\cos^2{\xi_l(d)\sec^2\overline{\varphi}}})}\right)\cos\overline{\varphi}d \overline{\varphi}},
            \end{aligned}
        \end{equation}
        When the distance $d$ exceeds the maximum coverage boundary $r_{l,\max}$, indicating no visible satellites within the serving area of layer $l$, the corresponding probability is given by
        \begin{equation}\label{dis3}
            \begin{aligned}
                \mathrm{P}_l^{\mathrm{nosat}}&=\Pr\left(D_l>d\right)\\
             &=e^{-\lambda_l\int_0^{\xi_{\beta,l}}\left(1-e^{-\frac{\mu_1}{\pi}\arcsin(\sqrt{1-\cos^2{\xi_{\beta,l}\sec^2\overline{\varphi}}})}\right)\cos\overline{\varphi}d \overline{\varphi}},
            \end{aligned}
        \end{equation}
        where $\xi_{\beta,l}= \arccos(\frac{r_e^2+r_l^2-r_{l,\max}^2}{2r_er_l})$. Based on the above analysis, the CCDF of $D_l$ can be expressed as
        \begin{equation}\label{dis4}
            \begin{aligned}
             &F'_{D_l}(d)=\Pr\left(D_l>d\right)\\
             &=\begin{cases}
                 1 & d<r_{l,\min}\\
                 \exp(-\lambda_l\int_0^{\xi_l(d)}(1-\\
                 e^{-\frac{\mu_l}{\pi}\arcsin(\sqrt{1-\cos^2{\xi_l(d)\sec^2\overline{\varphi}}})})
                 \cos\overline{\varphi}d \overline{\varphi})
                  & \begin{aligned}
                      &r_{l,\min}<d\\
                      &<r_{l,\max}
                  \end{aligned} \\
                 \mathrm{P}_l^{\mathrm{nosat}} & d>r_{l,\max} .
             \end{cases}        
            \end{aligned}
        \end{equation}

    To obtain the connectivity probability in (\ref{al2}), we should derive the PDF of the distance distribution $D_l$. Based on (\ref{dis4}), the PDF of distance distribution $D_l$ can be derived as 
    \begin{equation}\label{dstpdf}
        \begin{aligned}
             f_{D_l}(d)& = \frac{\partial}{\partial d}\Pr\left(D\leq d\right)
             = \frac{\partial}{\partial d}(1-F_{D_l}(d))\\
             &=\frac{\partial}{\partial d}\left(\exp\left(-\lambda_l\int_0^{\xi_l(d)}(1-
             \right.\right.\\
             &\left.\left.e^{-\frac{\mu_l}{\pi}\arcsin\left(\sqrt{1-\cos^2{\xi_l(d)\sec^2\overline{\varphi}}}\right)}\right)
                 \cos\overline{\varphi}d \overline{\varphi}\right),           
        \end{aligned}
    \end{equation}
    According to the chain rule and Leibniz integral rule, (\ref{dstpdf}) can be calculated as
        \begin{equation}\label{dstpdf1}
        \begin{aligned}
             f_{D_l}(d)& = \lambda_l F_{D_l}(d)g_l(d),          
        \end{aligned}
    \end{equation}
    where $g_l(d)$ is expressed as
            \begin{equation}\label{dstpdf2}
        \begin{aligned}
             g_l(d)=\frac{\mu_l d}{\pi r_e r_l}\int_0^{\xi_l(d)}\frac{e^{-\frac{\mu_l}{\pi}\arcsin\left(\sqrt{1-\cos^2{\xi_l(d)\sec^2\overline{\varphi}}}\right)}}{\sqrt{1-\cos^2{\xi_l(d)}\sec^2\overline{\varphi}}} d\overline{\varphi}.    
        \end{aligned}
    \end{equation}
    By substituting (\ref{dis4}) and (\ref{dstpdf1}) into (\ref{al2}), the connectivity probability $\Pr(\mathcal{A}_l)$ can be further calculated as
    \begin{equation}\label{al3}
             \Pr(\mathcal{A}_l)
             = \int_{r_{l,\min}}^{r_{l,\max}}\lambda_lg_l(v)F_{D_l}(v)\left(\prod_{\substack{m=1\\ m\neq l}}^{L}\Pr(D_m>\sqrt{\tau_{m,l}}v)\right)dv.
    \end{equation}

    \subsection{Coverage Analysis}
 
\begin{figure}
        \centering
        \includegraphics [width=0.48\textwidth] {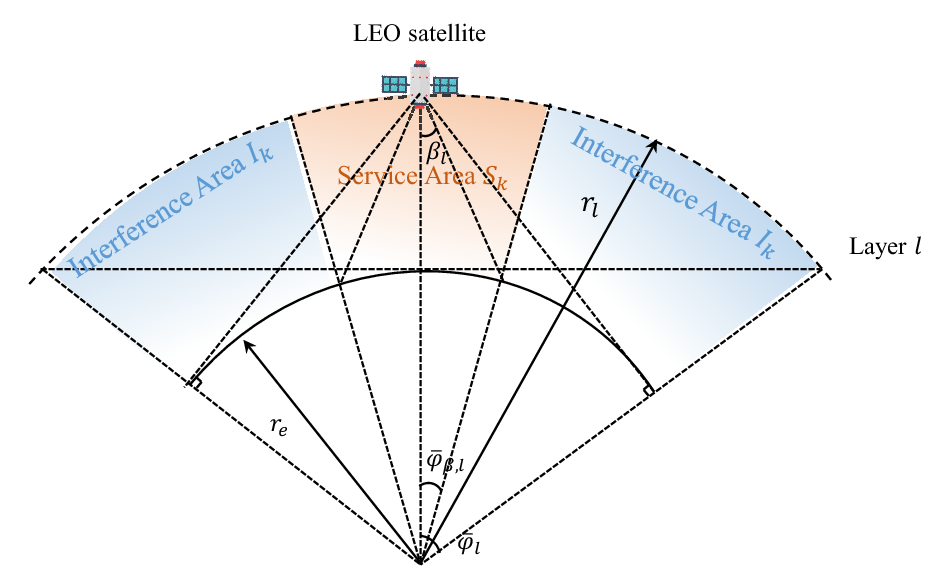}
        \caption {Coverage model of a LEO satellite. }
        \label{cov_model}
    \end{figure}

    Then, we analyze the coverage probability conditionally on the connected satellite of layer $l$ $\Pr(\gamma_{\star}>\gamma_{\mathrm{th}}|\mathcal{A}_l)$. For simplicity, we omit the subscript of $d_{l,m}$ as $d$. With (\ref{cdfh}), the coverage probability can be further obtained as 
    \begin{equation}\label{covgama2}
        \begin{aligned}
             \mathcal{F}'_{\gamma_\star|\mathcal{A}_l}(\gamma_{\mathrm{th}}) = &\mathbb{E}_{d,\mathcal{Z},{Z_{\star}}}\Big[Q_1\Big(\sqrt{2K}, \\
             &\sqrt{\frac{2(1+K)(I^\star+\sigma_0^2)\gamma_{\mathrm{th}}}{\Omega_s g_t p_l G_{\mathrm{sat},1} d^{-2}}}\Big)\Big],\\
        \end{aligned}        
    \end{equation}
    where $I^\star$ denotes the interference signal power received by the device $X_U$, which can be expressed as
    \begin{equation}\label{Itot}
        I^\star = \sum_{k=1}^L I_{k},
    \end{equation}
    where $I_{k}$ denotes the interference from the satellites of layer $k$. Specifically, the interference $I_{k}$ is composed of two parts, the signal from the interference area $I_k^{\mathrm{Inf}}$ and the service area $I_k^{\mathrm{Sev}}$. As illustrated in Fig. \ref{cov_model}, $\mathcal{I}_k$ is the interference area of layer $k$, which contains satellites that exclusively contribute to interference and $\mathcal{S}_k$ is the serving area where IoT devices may establish connections with satellites. This spatial decomposition enables precise evaluation of both desired signal and interference components. In other words, $\mathcal{S}_k$ is defined by the maximum coverage angle $\beta_l$ and $\mathcal{I}_k$ comprises all complementary satellite positions within the orbital plane. Such interference analysis maintains full consideration of antenna patterns, propagation losses, atmospheric effects, and other potential factors that may influence transmission performance. In this context, $I_k^{\mathrm{Inf}}$ and $I_k^{\mathrm{Sev}}$ can be expressed respectively as
       \begin{equation}\label{Isev}
       \begin{aligned}
                    I_k^{\mathrm{Sev}}&= \sum_{m'\in\mathcal{S}_k} p_k 
            g_{k,m'}^2|h_{k,m'}|^2\\
            &=g_t p_kG_{\mathrm{sat},k} \sum_{m'\in\mathcal{S}_k}|h_{k,m'}|^2 d_{k,m'}^{-2},
        \end{aligned}
    \end{equation}
  and
    \begin{equation}\label{Iinf}
        \begin{aligned}
                    I_k^{\mathrm{Inf}}& = \sum_{m'\in\mathcal{I}_k} p_k 
            g_{k,m'}^2|h_{k,m'}|^2
            =g_t p_k\sum_{m'\in\mathcal{I}_k} |h_{k,m'}|^2 d_{k,m'}^{-2}.
        \end{aligned}
    \end{equation}
    Noticeably, it is difficult to calculate the expectation in (12) due to the complex expression of Marcum Q-function, which motivates us to use the finite General Dirichlet series to approximate it \cite{fds}. Specifically, the CCDF of Rician distributed $|h_{l,m}|^2$ can be approximated as
    \begin{equation}\label{Q1approx}
         \mathcal{F}'_{|h_{l,m}|^2}(x) \approx\sum_{n=1}^{N}\chi_n\exp(-\eta_n x)=\widetilde{\mathcal{F}}_{|h_{l,m}|^2}'(x),
    \end{equation}
    where the value of $N$ is the number of General Dirichlet series terms where a large number of $N$ yields a higher accuracy of approximation. $\chi_n$ and $\eta_n$ are parameters, such that $\eta_n>0$. Moreover, we obtain $\sum_{n=1}^{N}\chi_n =1$ from $\mathcal{F}'_{|h_{l,m}|^2}(0) = \widetilde{\mathcal{F}}'_{|h_{l,m}|^2}(0)$. To evaluate the approximation error, we define the approximation error $\mathcal{E} (K)$ function as 
    \begin{equation}\label{apperror}
        \begin{aligned}
         \mathcal{E} (K)& =  \int_0^{\infty}\Big( \mathcal{F}'_{|h_{l,m}|^2}(x)-\widetilde{\mathcal{F}}_{|h_{l,m}|^2}'(x)\Big)^2 dx.
         \end{aligned}   
    \end{equation}
    It can be found in (\ref{apperror}) that the integration is difficult to be calculated, and thus we adopt a numerical approach that the approximation error is expressed as 
    \begin{equation}\label{apperror2}
        \begin{aligned}
         \widehat{\mathcal{E}} (K)
         = \delta \sum_{i=0}^{\infty}\Big( \mathcal{F}'_{|h_{l,m}|^2}(i\delta)-\widetilde{\mathcal{F}}_{|h_{l,m}|^2}'(i\delta)\Big)^2,
         \end{aligned}   
    \end{equation}
    where $\delta$ is a small number and $\widehat{\mathcal{E}} (K)\rightarrow\mathcal{E} (K)$ as $\delta\rightarrow0$. Based on (\ref{apperror2}), the appropriate parameters $\chi_n$ and $\eta_n$ are determined by solving the following optimization problem formulated as
    \begin{subequations}
   	\begin{eqnarray}
    Q1:~ \underset{\chi_n,\eta_n}{\mathop{\text{minimize}}}\!\!&&\!\!\!\!\!\!\widehat{\mathcal{E}} (K)\label{OP1obj}\\
   		\textrm{s.t.}&&\!\!\!\!\!\!  \sum_{n=1}^{N}\chi_n = 1, \label{OP1st1}\\
      &&\!\!\!\!\!\! \eta_n > 0.\quad \forall n\in \{1, 2, \cdots, 
 N\}\label{OP1st2}                .
   	\end{eqnarray}
   \end{subequations}
     It is obvious that the optimization problem $Q1$ is a non-convex problem, for which obtaining the global optimization solution in polynomial time seems to be impossible. Fortunately, intelligent optimization techniques can be applied to solve such problems. Specifically, we utilize the genetic algorithm (GA), a classical intelligent optimization method to solve the problems under several values of $K$. Given that the objective function (\ref{OP1obj}) involves an infinite summation, its exact computation is prohibitively resource-intensive while contributing minimally to accuracy improvements. Recognizing that the Marcum Q-function $Q_1(a,b)$ decays exponentially with $b$, we truncate the summation by discarding terms beyond values of $x$ upper than several $x_{\max}$ in order to facilitate optimization. Thus, the upper limit of the summation is replaced by $i_{\max} = x_{\max}/\delta$. For parameter settings, we set $\delta = 10^{-4}$, $N=4$, $\Omega_s = 1$ and $x_{\max}= 10$. In general, the Rician factor $K$  is $1\leq K\leq 10$ \cite{rik}. {Specifically, the effectiveness and accuracy of the proposed approximation method is given out in Section V. A.}
%
    
    
    With the proposed approximation method (\ref{Q1approx}), (\ref{covgama2}) can be further expressed as
    \begin{equation}\label{covgama3}
        \begin{aligned}
             \mathcal{F}'_{\gamma_m}&(\gamma_{\mathrm{th}}) =\mathbb{E}_{d,\mathcal{Z},{Z_{\star}}}\Big[\sum_{n=1}^N\chi_n e^{-\eta_n\frac{(I^\star+\sigma_0^2)\gamma_{\mathrm{th}}}{g_t p_l G_{\mathrm{sat},l} d^{-2}} }\Big] \\
             & = \sum_{n=1}^N\chi_n\cdot\mathbb{E}_{d,\mathcal{Z},{Z_{\star}}}\Big[e^{-\frac{\eta_n\sigma_0^2\gamma_{\mathrm{th}}d^{2}}{g_t p_l G_{\mathrm{sat},l} }}\mathcal{L}_{I^\star}(\frac{\eta_n\gamma_{\mathrm{th}}d^{2}}{g_t p_l G_{\mathrm{sat},l} })\Big],
        \end{aligned}        
    \end{equation}
    where $\mathcal{L}_{I^\star}(s)$ represents the Laplace transform of the total interference $I^\star$ evaluated at $s$ conditionally on $d,\mathcal{Z}$ and ${Z_{\star}}$. Based on (\ref{Itot}), $\mathcal{L}_{I^\star}(s)$ can be further decomposed according to the corresponding layers, which is expressed as
    \begin{equation}\label{lplce}
             \mathcal{L}_{I^\star}(s) = \prod_{k=1}^{L}\mathcal{L}_{I_k}(s),  
    \end{equation}
    where $\mathcal{L}_{I_k}(s)$ denotes the Laplace transform of the interference of layer $k$. By making use of the properties of CPP and HPPP, we can obtain the following proposition:
    
    \textbf{Proposition 1:} The Laplace transform of the interference of layer $k$ $\mathcal{L}_{I_{k}}(s)$ can be calculated as (\ref{lpm}) on the top of the next page.
    \begin{figure*}
      \begin{equation}\label{lpm}
   \begin{aligned}
    	  \mathcal{L}_{{I_k}} (s)&=\prod_{Z_i\in \mathcal{Z}_k}^{|\varphi_i-\pi/2|\leq \xi_k}e^{-\frac{\mu_k}{\pi}\left(\int_{w_{k,\varphi_i,1}}^{w_{k,\varphi_i,2}}
            1-\mathcal{L}_{|h|^2}\left(\frac{sg_tp_kG_{\mathrm{sat},k}}{f_{k,\varphi_i}^2(w)}\right)d w+\int_{w_{k,\varphi_i,2}}^{w_{k,\varphi_i,3}}
            1-\mathcal{L}_{|h|^2}\left(\frac{sg_tp_k}{f_{k,\varphi_i}^2(w)}\right)d w\right)}\\
            &\cdot \prod_{Z_i\in\mathcal{Z}_k}^{\xi_k\leq|\varphi_i-\pi/2|\leq \overline{\varphi}_{\beta, k}}e^{-\frac{\mu_k}{\pi}\left(\int_{0}^{w_{k,\varphi_i,2}}
            1-\mathcal{L}_{|h|^2}\left(\frac{sg_tp_kG_{\mathrm{sat},k}}{f_{k,\varphi_i}^2(w)}\right)d w+\int_{w_{k,\varphi_i,2}}^{w_{k,\varphi_i,3}}
            1-\mathcal{L}_{|h|^2}\left(\frac{sg_tp_k}{f_{k,\varphi_i}^2(w)}\right)d w\right)}\\
            &\prod_{Z_i\in \mathcal{Z}_k}^{\overline{\varphi}_{\beta,k}\leq |\varphi_i-\pi/2|\leq \overline{\varphi}_k}e^{-\frac{\mu_k}{\pi}\int_{0}^{w_{k,\varphi_i,3}}1-\mathcal{L}_{|h|^2}\left(\frac{sg_tp_k}{f_{k,\varphi_i}^2(w)}\right)d w}.
    \end{aligned}
   \end{equation}
   \hrulefill
   \end{figure*}
   \begin{IEEEproof}
       See Appendix A.
    \end{IEEEproof}

   By using the Fubini’s theorem, we first unfold the expectation of $d$ in (\ref{covgama3}). Thereby, the coverage probability can be further derived as
     \begin{equation}\label{cov4}
        \begin{aligned}
             \mathcal{F}'_{\gamma_\star|\mathcal{A}_l}(\gamma_{\mathrm{th}})
             &=\sum_{n=1}^N\chi_n\cdot \int_{r_{l,\min}}^{r_{l,\max}}e^{-\frac{\eta_n\sigma_0^2\gamma_{\mathrm{th}}r^{2}}{g_t p_l G_{\mathrm{sat},l}}}\cdot \\
             &\mathbb{E}_{\mathcal{Z}}\left[\mathbb{E}_{Z_{\star}}\left[\mathcal{L}_{I_l^\star}\left(\frac{\eta_n\gamma_{\mathrm{th}}r^2}{g_t p_l G_{\mathrm{sat},l} }\right)f_{D_l|Z_{\star},\mathcal{Z}}(r)\right]\right]dr,
        \end{aligned}        
    \end{equation}
    where $f_{D_l|Z_{\star},\mathcal{Z}}(r)$ denotes the PDF of $D_l$ conditionally on $Z_{\star}$ and $\mathcal{Z}$, which can be derived as
    \begin{equation}\label{dstpdf2}
        \begin{aligned}
             f_{D_l|Z_{\star},\mathcal{Z}}&(d)=  \partial_d\Pr\left(D\leq d|\mathcal{Z},Z_{\star}\right)\\
             = &\partial_d\left(1-\Pr\left(D_j>d|X_j\in \psi_\star\right)\right)\\
             &\cdot\left(\Pr\left(D_j>d|X_j\in \psi_i,Z_i\in\mathcal{Z}_l\right)\right)\\
             &\cdot\prod_{k\neq l}\left(\Pr\left(D_j>d|X_j\in \psi_i,Z_i\in\mathcal{Z}_k\right)\right)\\
           = &\frac{\mu_ld|\csc\varphi_{\star}|e^{-\frac{\mu_l}{\pi}\arcsin(\sqrt{1-\cos^2\xi_l(d)\csc^2{\varphi_{\star}}})}}{\pi r_lr_e\sqrt{1-\cos^2\xi_l(d)\csc^2\varphi_{\star}}}\cdot\\
            & \prod_{k=1}^L\prod_{Z_i\in \mathcal{Z}_k}^{|\varphi_i-\pi/2|<\xi_k}e^{-\frac{\mu_k}{\pi} \arcsin(\sqrt{1-\cos^2(\xi_k(d))\csc^2(\varphi_i)})}.
        \end{aligned}        
    \end{equation}

    By substituting (\ref{al2}), (\ref{lplce})-(\ref{dstpdf2}) in (\ref{covgama1}), we can obtain the expression for coverage probability by calculating the integration of $Z_{\star}$ and $\mathcal{Z}_l$, which can be expressed as (\ref{cov5}) on the top of the next page. Hence, the performance of the coverage probability can be evaluated directly. 
    \begin{figure*}
          \begin{equation}\label{cov5}
   \begin{aligned}
    	  \mathcal{F}'_{\gamma_\star}&(\gamma_{\mathrm{th}}) 
             =\sum_{l=1}^L\Pr{(\mathcal{A}_l)}\sum_{n=1}^N\chi_n\cdot \int_{r_{l,\min}}^{r_{l,\max}}\frac{\lambda_l\mu_lr}{\pi r_lr_e}e^{-\frac{\eta_n\sigma_0^2\gamma_{\mathrm{th}}r^{2}}{g_t p_l G_{\mathrm{sat},l}}}\\
             &\left(\int_{0}^{\xi_l(r)}\frac{e^{-\frac{\mu_l}{\pi}\arcsin(\sqrt{1-\cos^2{\xi_l(r)}\sec^2{v}})-\frac{\mu_l}{\pi}\left(\int_{w_{l,v,1}}^{w_{l,v,2}}
            1-\mathcal{L}_{|h|^2}\left(\frac{sg_tp_lG_{\mathrm{sat},l}}{f_{l,v}^2(w)}\right)d w+\int_{w_{l, v,2}}^{w_{l,v,3}}
            1-\mathcal{L}_{|h|^2}\left(\frac{sg_tp_1}{f_{l,v}^2(w)}\right)d w\right)}}{\sqrt{1-\cos^2{\xi_l(r)}\sec^2v}}dv\right)\\
            &\prod_{k=1}^L\exp{\left(-\lambda_k\int_0^{\xi_k(r)}1-e^{-\frac{\mu_k}{\pi}\left(\arcsin(\sqrt{1-\cos^2{\xi_k(r)}\sec^2{\overline{\varphi}}})+\int_{w_{k,\overline{\varphi},1}}^{w_{k,\overline{\varphi},2}}
            1-\mathcal{L}_{|h|^2}\left(\frac{sg_tp_kG_{\mathrm{sat},k}}{f_{k,\overline{\varphi}}^2(w)}\right)d w+\int_{w_{k,\overline{\varphi},2}}^{w_{k,\overline{\varphi},3}}
            1-\mathcal{L}_{|h|^2}\left(\frac{sg_tp_k}{f_{k,\varphi}^2(w)}\right)d w\right)}\cos{\overline{\varphi}}d\overline{\varphi}\right)}\\
            &\prod_{k=1}^L\exp{\left(-\lambda_k\int_{\xi_k(r)}^{\overline{\varphi}_{\beta,k}}1-e^{-\frac{\mu_k}{\pi}\left(\int_{0}^{w_{k,\overline{\varphi},2}}
            1-\mathcal{L}_{|h|^2}\left(\frac{sg_tp_kG_{\mathrm{sat},k}}{f_{k,\overline{\varphi}}^2(w)}\right)d w+\int_{w_{k,\overline{\varphi},2}}^{w_{k,\overline{\varphi},3}}
            1-\mathcal{L}_{|h|^2}\left(\frac{sg_tp_k}{f_{k,\overline{\varphi}}^2(w)}\right)d w\right)}\cos\overline{\varphi}d\overline{\varphi}\right)}\\
            &\prod_{k=1}^L\exp{\left(-\lambda_k\int_{\overline{\varphi}_{\beta, k}}^{\overline{\varphi}_{k}}1-e^{-\frac{\mu_k}{\pi}\left(\int_{0}^{w_{k,\overline{\varphi},3}}
            1-\mathcal{L}_{|h|^2}\left(\frac{sg_tp_k}{f_{k,\overline{\varphi}}^2(w)}\right)d w\right)}\cos{\overline{\varphi}}d\overline{\varphi}\right)}dr .        
    \end{aligned}
   \end{equation}
   \hrulefill
   \end{figure*}
    
   \subsection{Rate Analysis}
    Herein, we analyze the transmission rate of LEO satellite IoT constellations under two different transmission situations, i.e., infinite block-length packet transmission and finite block-length packet transmission. Firstly, we introduce the following lemma.

    \textbf{Lemma 1}: For a non-negative random variable $X$, the expectation of $X$ can be calculated as $\mathbb{E}[X]=\int_0^{\infty} \Pr{(X>x)}dx$.
    \begin{IEEEproof}
       We use $f(x)$ to denote the PDF of $X$. Then we have 
             \begin{equation}\label{ex}
        \begin{aligned}
             \mathbb{E}[X]&= \int_0^{\infty}yf(y)dy=\int_0^{\infty}\int_0^{x}f(y)dxdy\\
             &=\int_0^{\infty}\int_{x}^{\infty}f(y)dy dx = \int_0^{\infty} \Pr{(X>x)}dx.
        \end{aligned}        
    \end{equation}
       The proof is completed.
    \end{IEEEproof}
    Based on the above lemma, we analyze the transmission rate of LEO satellite IoT constellation under situations of infinite block length and finite block length, respectively.

    \emph{A) Infinite Block-Length Packet Transmission}: In the case of infinite block-length packet transmission, the transmission rate can be computed according to the shannon capacity, which is given by
        \begin{equation}\label{rate1}
         \mathcal{I}_m = \mathbb{E}_{\gamma_m}\left[\log_2{(1+\gamma_m)}\right].  
        \end{equation}
    Then, based on (\ref{cov5}) and Lemma 1, the transmission rate can be further calculated as
    \begin{equation}\label{rate2}
            \begin{aligned}
                \mathcal{I}_m& = \mathbb{E}_{\gamma_m}\left[\log_2{(1+\gamma_m)}\right]\\
                & = \int_0^{\infty}{\Pr{(\log_2(1+\gamma_m)>t)}} dt\\
                & = \int_0^{\infty}{\Pr{(\gamma_m>2^t-1)}} dt.\\
            \end{aligned} 
        \end{equation}

    \emph{B) Finite Block-Length Packet Transmission}: In the context of LEO satellite IoT, devices typically operate under finite block-length transmission regimes due to the short-packet nature of IoT traffic. Within this framework, communication is organized in discrete time slots, with each transmission block comprising $M$ channel uses, defining the block length. Unlike conventional scenarios where Shannon capacity provides the fundamental limit, the finite block-length regime requires alternative performance metrics. Following the approach in \cite{fbl}, we adopt a more accurate approximation of the achievable rate that properly captures the fundamental trade-offs in short-packet communications. Particularly, given a target error probability $\delta_{\mathrm{th}}$, the approximated transmission rate can be expressed as
        \begin{equation}\label{rate1}
         \mathcal{I}_m^{\mathrm{FBL}} = \mathbb{E}_{\gamma_m}\left[\log_2{(1+\gamma_m)-\sqrt{\frac{V(\gamma_m)}{M}}\frac{Q^{-1}(\delta_{\mathrm{th}})}{\ln 2}}\right], 
        \end{equation}
    where $Q^{-1}(\cdot)$ refers to the inverse standard Gaussian $Q$ function, $V(\gamma_m)= \left(1-\left(1+\gamma_m\right)^{-2}\right)(\log_2e)^2$ stands for channel dispersion, which measures the stochastic channel variability with respect to a deterministic channel with the same channel capacity. Then, based on the (\ref{cov5}) and the Lemma above, the transmission rate can be  further calculated as
    \begin{equation}\label{rate2}
            \begin{aligned}
                \mathcal{I}&_m^{\mathrm{FBL}} =  \mathbb{E}_{\gamma_m}\left[\log_2{(1+\gamma_m)}\right]-\frac{Q^{-1}(\delta_{\mathrm{th}})}{\ln{2}\sqrt{M}}\mathbb{E}_{\gamma_m}[\sqrt{V(\gamma_m)}]\\
                 = &\mathcal{I}_m-\frac{Q^{-1}(\delta_{\mathrm{th}})}{\ln{2}\sqrt{M}}\cdot\\
                &\int_0^{\infty}{\Pr{\left(\sqrt{\left(1-\left(1+\gamma_m\right)^{-2}\right)(\log_2e)^2}>t\right)}}dt\\
                 = &\mathcal{I}_m-\frac{Q^{-1}(\delta_{\mathrm{th}})}{\ln{2}\sqrt{M}}\cdot
                \int_0^{\log_2 e}{\Pr{\left(\gamma_m>\frac{\log_2 e}{\sqrt{(\log_2 e)^2-t^2}}-1\right)}}dt\\
                =& \mathcal{I}_m--\frac{Q^{-1}(\delta_{\mathrm{th}})}{\ln{2}\sqrt{M}}\int_0^{\log_2 e}\mathcal{F}'_{\gamma_m}(\frac{\log_2 e}{\sqrt{(\log_2 e)^2-t^2}}-1)dt.
            \end{aligned}  
        \end{equation}

    Now, we have obtained the expressions for transmission rate in the cases of infinite block-length packet transmission and finite block-length packet transmission. Hence, we can evaluate the performance of transmission rate according to the parameters of LEO satellite IoT constellation.

\section{Simulation Results}

    In this section, we provide extensive simulation results to validate the accuracy of the derived expressions for connectivity probability, coverage probability, and transmission rate via the Monte Carlo simulation method with $10^6$ number of iterations to ensure precision. The main parameters of simulations are set according to TR 38.821 \cite{tr}, which are shown in Table \ref{Simulation}\footnote{The proposed analytical framework can be extended to $N$-layer ($N > 3$) LEO satellite IoT constellations, but numerical experiments showed that increasing the number of layers introduces heightened sensitivity to system parameters, e.g., constellation scale and orbital altitude, potentially obscuring critical system insights. To maintain analytical tractability while ensuring meaningful interpretation of results, we focus our simulation study on a two-layer LEO satellite IoT constellation architecture.}. Specifically, we consider a constellation consisting of two-layer satellites, where satellites in layer $1$ have smaller orbit radius than satellites in layer $2$. Via simulations, we will try to reveal some insights by comparing with the classical constellations. 

    \begin{table}
	\small
	\centering
	\caption{Main System Parameters For LEO Satellite IoT Constellation}\label{Simulation}
	\begin{tabular}{|c|c|}
		\hline
		Parameter & Value\\\hline\hline
        Earth's radius $r_e$ & 6378.14 km \\\hline
		Carrier frequency $f$& 5 GHz  \\\hline
		Carrier bandwidth $B$  & 250 MHz  \\\hline
		  IoT device antenna gain $G_{\mathrm{dev}}$  &  3 dBi  \\\hline
		  Rain fading mean  $\mu_r$& -2.6 dB \\\hline
            Rain fading variance  $\sigma_r^2$& 1.63 dB \\\hline
		Rician factor $K$ & 10 dB \\\hline
		Noise variance  $\sigma_0^2$ & $-150$ dBm\\ \hline
            Number of layers $L$ & 2\\ \hline
            Orbit radius $r_l$  & [6800 km, 7400 km] \\ \hline
            Satellite antenna gain $G_{\mathrm{sat},l}$  & 20 dBi  \\\hline
            Average transmit power $\xi_l$ & [30 dBm, 30 dBm]\\ \hline
		Satellite coverage angle  $\beta_l$ & $[60^\circ,60^\circ]$ \\ \hline
            Satellite density  $\mu_l$ & $[20 ,20]$ \\ \hline
            Orbit density  $\lambda_l$ & $[20 ,20]$ \\ \hline
            SINR threshold $\gamma_{th}$ & -10 dB \\ \hline
            Target error rate $\delta_{th}$ &$10^{-4}$ \\ \hline
            Number of time intervals $M$ & $10^{4}$ \\ \hline
	\end{tabular}
\end{table}	
    
    \begin{figure}[h]
        \centering
        \includegraphics [width=0.45\textwidth] {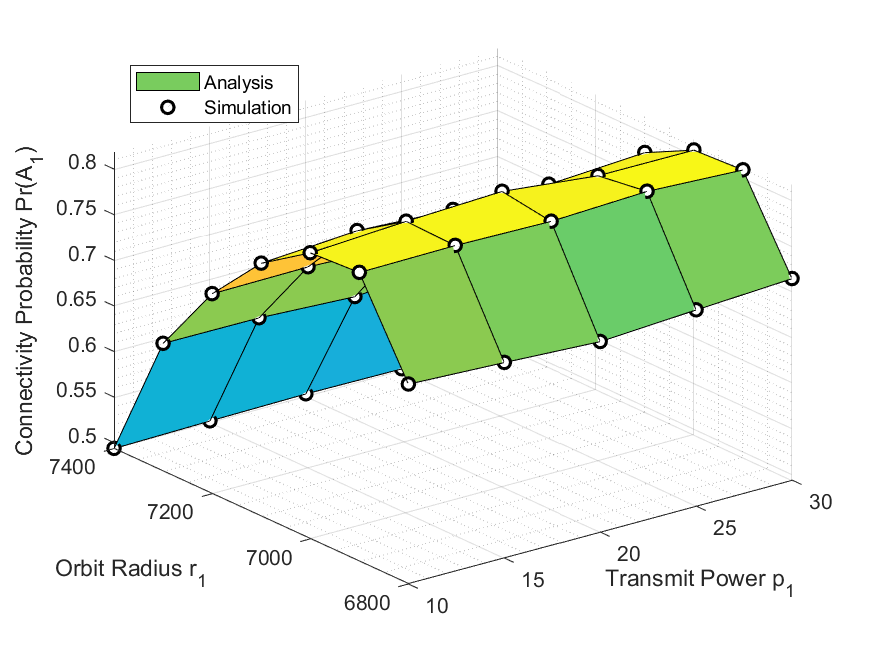}
        \caption {Connectivity probability $\Pr{(\mathcal{A}_1)}$ performance under various orbit radius $r_1$ and transmit power $p_1$.}
        \label{conprbr1}
    \end{figure}

    \begin{figure}[h]
        \centering
        \includegraphics [width=0.4\textwidth] {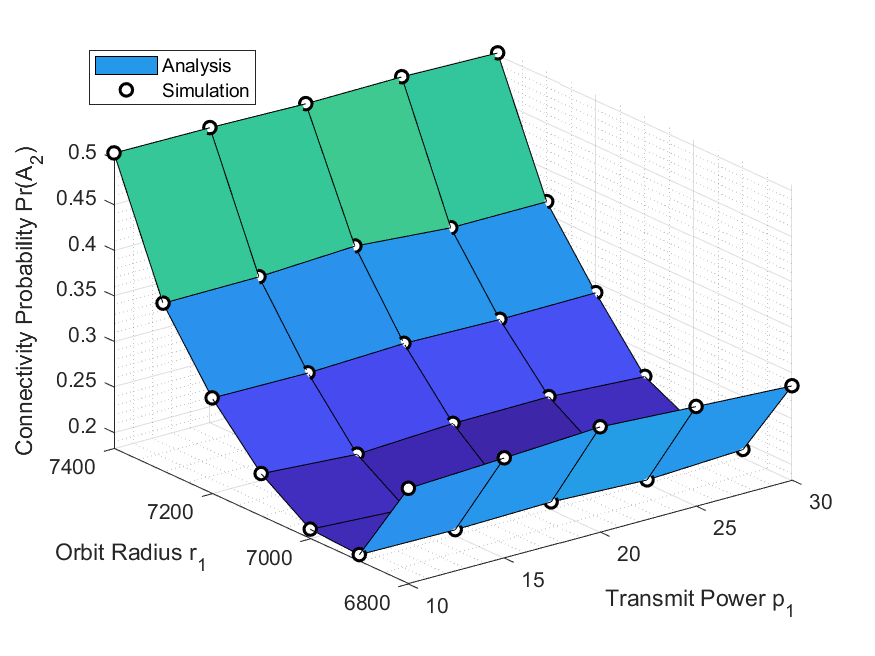}
        \caption {Connectivity probability $\Pr{(\mathcal{A}_2)}$ performance under various orbit radius $r_1$ and transmit power $p_1$.}
        \label{conprbr2}
    \end{figure}

    In Fig. \ref{conprbr1} and Fig. \ref{conprbr2}, we plot the connectivity probability of two layers $\Pr{(\mathcal{A}_1)}$ and $\Pr{(\mathcal{A}_2)}$ under various orbit radius $r_1$ and transmit power $p_1$. As expected, the analytical results closely align with the simulations, confirming the accuracy of the theoretical model. Notably, for orbital radius between 6900 km and 7400 km, $\Pr{(\mathcal{A}_1)}$ decreases while $\Pr{(\mathcal{A}_2)}$ increases with $r_1$. This trend arises because larger orbital radius leads to greater path loss, reducing received signal power and increasing the likelihood of IoT devices connecting to satellites in other layers. Conversely, for $r_1$ between 6800 km and 6900 km, $\Pr{(\mathcal{A}_1)}$ increases while $\Pr{(\mathcal{A}_2)}$ decreases. In this regime, the reduced orbital radius shrinks the satellite coverage area under fixed coverage angles $\beta_l$ and number of satellites. While lower orbits enhance signal strength, the diminished coverage area reduces the probability of satellites being within a device’s communication range, favoring connections with higher-altitude satellites. Therefore, when designing a LEO satellite IoT constellation, we should select an appropriate orbital radius based on the actual satellite scale to improve communication quality. Furthermore, at the same orbital radius, $\Pr{(\mathcal{A}_1)}$ and $\Pr{(\mathcal{A}_2)}$ remain stable across transmit power variations. The observed stability stems from the fact that in densely populated LEO satellite IoT constellations, connectivity probabilities are primarily governed by inter-satellite interference rather than noise or transmit power levels.

    \begin{figure}
        \centering
        \includegraphics [width=0.4\textwidth] {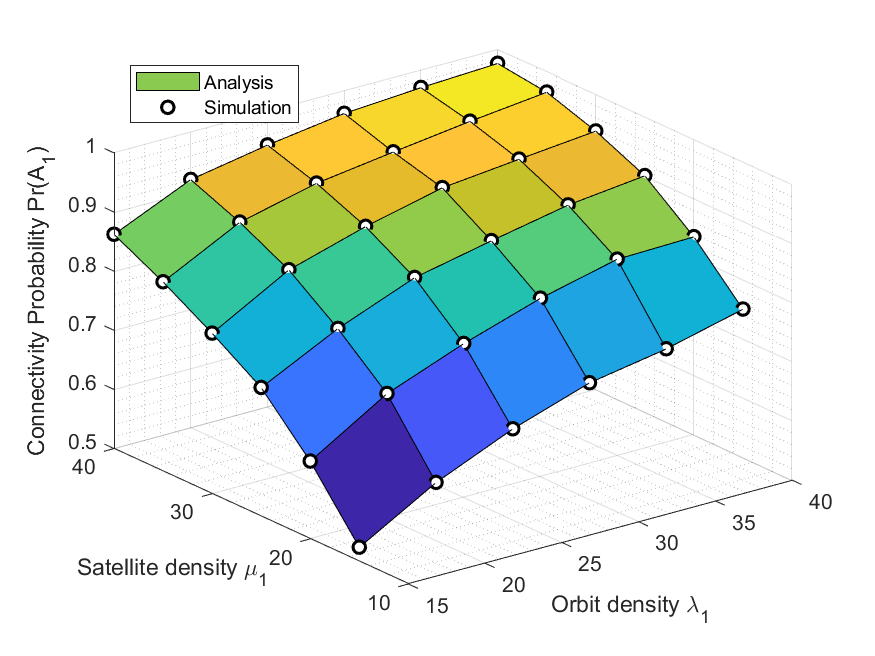}
        \caption {connectivity probability $\Pr{(\mathcal{A}_1)}$ performance under various orbit densities  $\lambda_1$ and satellite densities $\mu_1$.}
        \label{conprb11}
    \end{figure}

    \begin{figure}
        \centering
        \includegraphics [width=0.4\textwidth] {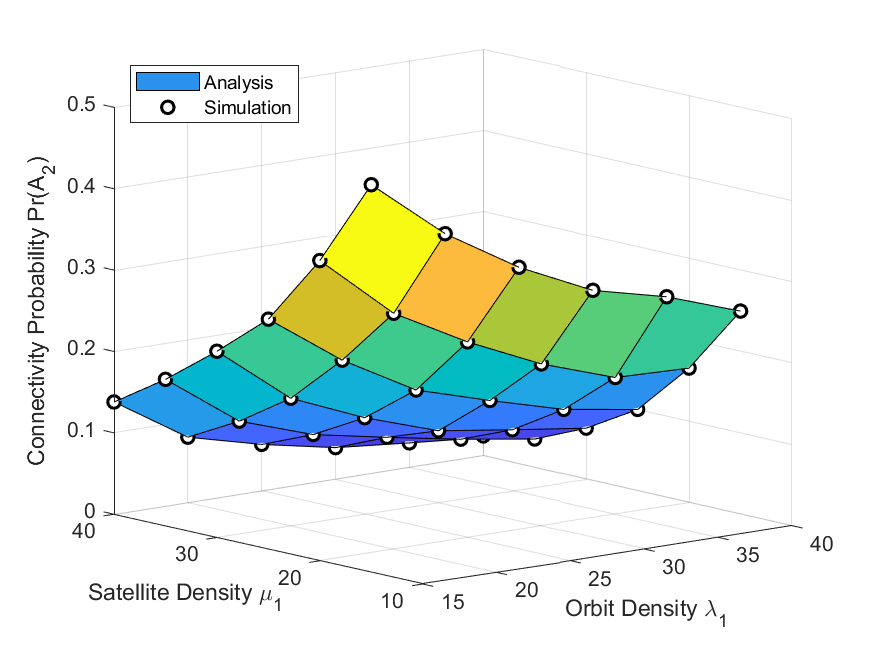}
        \caption {connectivity probability $\Pr{(\mathcal{A}_2)}$ performance under various orbit densities  $\lambda_1$ and satellite densities $\mu_1$.}
        \label{conprb12}
    \end{figure}

    The number of satellites per orbit and the number of orbits are two important indicators of the LEO satellite IoT constellation. Therefore, we investigate the impacts of these two parameters on the connectivity probabilities $\Pr{(\mathcal{A}_1)}$ and $\Pr{(\mathcal{A}_2)}$. As shown in Figs. \ref{conprb11} and \ref{conprb12}, increasing satellite density and orbital density in layer $1$ leads to a rise in $\Pr{(\mathcal{A}_1)}$ and a corresponding decrease in $\Pr{(\mathcal{A}_2)}$. This behavior occurs because these two factors collectively determine the constellation scale in layer $1$. When both parameters increase, IoT devices exhibit a higher probability of establishing connections with satellites of layer $1$. This suggests an inherent trade-off in multiple-layer constellation design, where enhancing connectivity in one layer necessarily reduces the connectivity probability in others, highlighting the importance of balanced resource allocation across orbital layers.
    
    \begin{figure}
        \centering
        \includegraphics [width=0.45\textwidth] {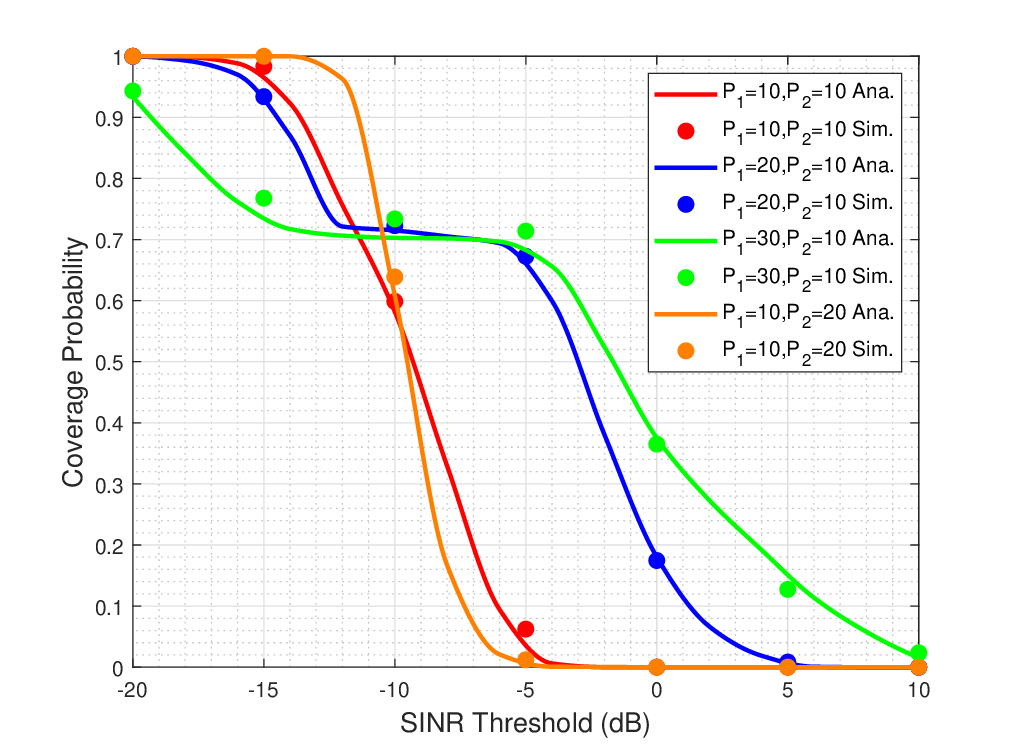}
        \caption {Coverage probability performance under different transmit power versus different SINR thresholds.}
        \label{cop_pow}
    \end{figure}
    Next, we examine the performance of coverage probability for the proposed model in Fig. \ref{cop_pow}. As expected, the analytical results closely align with the simulations, confirming the accuracy of the theoretical analysis. As the transmit power of satellites in layer 1 $p_1$ increases, the decline in coverage probability becomes more gradual. This trend occurs because a larger $p_1$ improves the SINR for connections to satellites of layer $1$.  Notably, when $p_1$ is set to 20 dBm or 30 dBm, the coverage probability stabilizes near 0.7 for certain SINR thresholds, indicating that devices predominantly connect to layer 1 satellites under these conditions —consistent with the coverage probability definition and the simulation results in Fig. \ref{conprbr1} and Fig. \ref{conprbr2}. Conversely, variations in $p_2$ induce more significant fluctuations in coverage probability, suggesting that the transmit power adjustments of layer $2$ have a stronger impact on system performance. These findings underscore a hierarchical power-dependency in multiple-layer constellations: while layer 1 power governs baseline connectivity stability, layer 2 power serves as a critical tuning parameter for coverage optimization. Thus, constellation designs must strategically balance inter-layer power allocation to meet specific SINR and coverage requirements.

    \begin{figure}[h]
        \centering
        \includegraphics [width=0.5\textwidth] {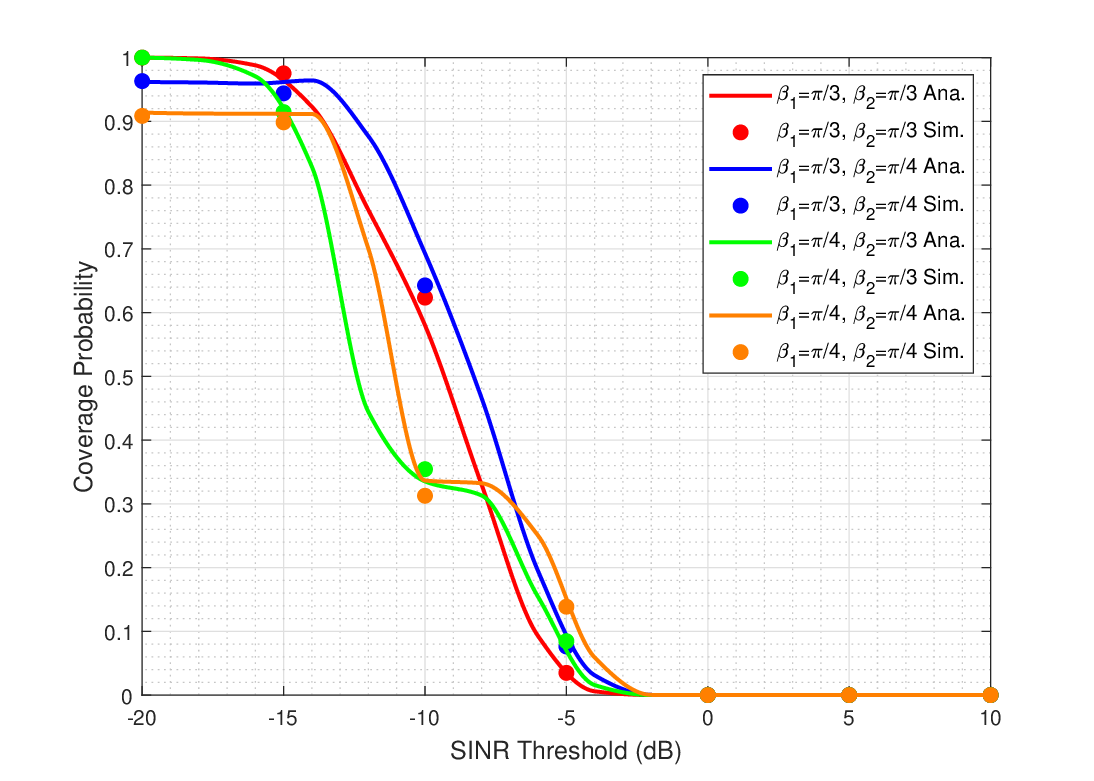}
        \caption {Coverage probability performance under different antenna coverage angles versus different SINR thresholds.}
        \label{cop_theta}
    \end{figure}

    Fig. \ref{cop_theta} shows the performance of coverage probability under different SINR thresholds for various combinations of satellite coverage angle $\beta_l$ of each layer. Specifically, when the satellite coverage angle of layer 1 $\beta_1$ decreases, there is a distinct segmentation phenomenon with the increasing SINR thresholds. For a given SINR threshold (e.g., -8 dB), this segmentation indicates communication occurring primarily with satellites in layer $1$. This is because reduced $\beta_1$ value diminishes the coverage range of these satellites, yielding a connectivity probability of approximately 0.31 with the IoT devices for satellites in layer $1$. However, the reduction in $\beta_2$ affects coverage differently. Under low SINR thresholds, the coverage probability declines because the satellites in layer $2$, e.g., higher-orbit satellites with smaller coverage angles $\beta_2$, provide less overlapping coverage area. This reduction in effective coverage increases the probability of no visible satellites $\mathrm{P}_l^{\mathrm{nosat}}$ mentioned in (\ref{dis3}) throughout the constellation. These findings reveal a critical design constraint in LEO satellite IoT constellation, while decreasing coverage angles $\beta_l$ may improve signal quality for connected devices, it simultaneously reduces overall network availability. This fundamental trade-off necessitates careful optimization of satellite coverage angles to balance connection reliability within service area coverage in multiple-layer constellations.

    \begin{figure}[h]
        \centering
        \includegraphics [width=0.4\textwidth] {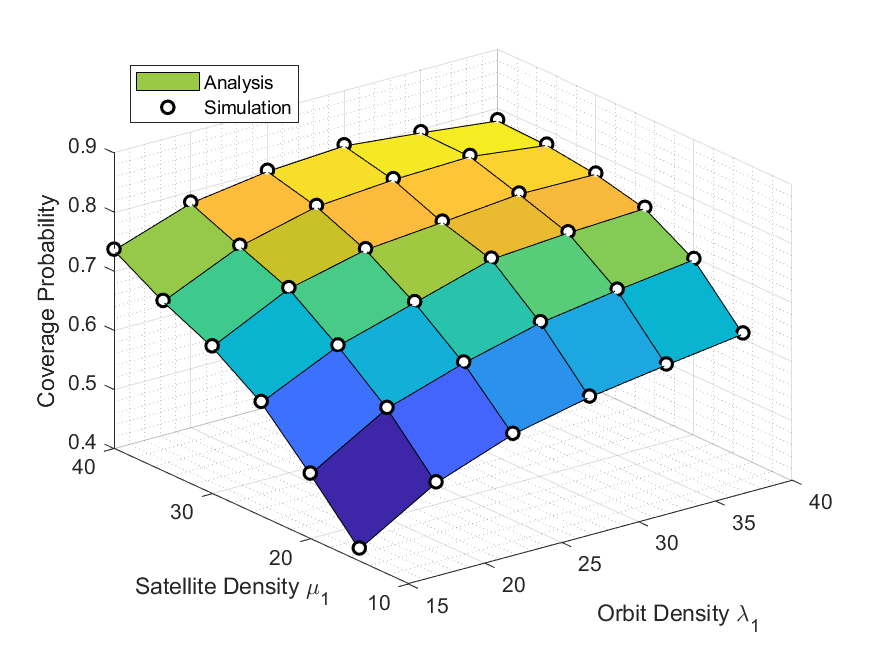}
        \caption {Coverage probability performance under different satellite densities $\mu_1$ and orbit densities $\lambda_1$ of layer $1$.}
        \label{covprb1}
    \end{figure}

    \begin{figure}[h]
        \centering
        \includegraphics [width=0.4\textwidth] {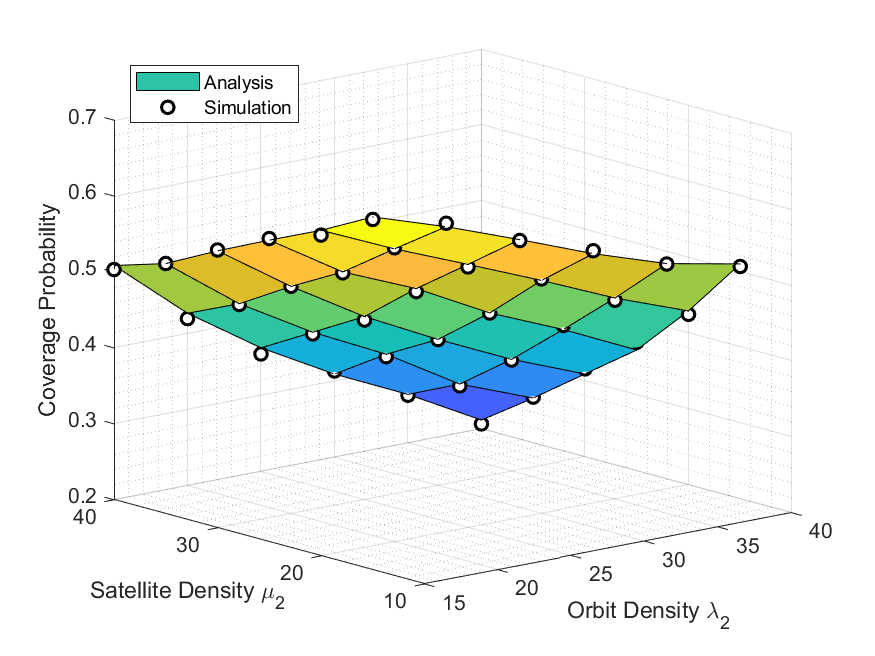}
        \caption {Coverage probability performance under different satellite densities $\mu_2$ and orbit densities $\lambda_2$ of layer $2$.}
        \label{covprb2}
    \end{figure}

   Fig. \ref{covprb1} and Fig. \ref{covprb2} investigate the effects of satellite densities and orbit densities on coverage probability in LEO satellite IoT constellation. Fig. \ref{covprb1} reveals a non-monotonic relationship, where the coverage probability initially rises before declining as the satellite density $\mu_1$ and orbit density $\lambda_1$ increase. This is because higher densities first improve connectivity probabilities with LEO satellites in layer $1$, whose conditional coverage probability exceeds that of layer 2 satellites. However, beyond a critical density threshold, growing interference from additional satellites degrades overall coverage performance.
   Moreover, Fig. \ref{covprb2} demonstrates a contrasting monotonic decrease in coverage probability with increasing densities for satellites in layer $2$. As a higher orbit altitude with inherently lower connectivity probability, the coverage performance of satellites in layer 2 is more sensitive to density-induced interference effects without benefiting from the initial positive network effects observed in layer 1. Therefore, it highlights a fundamental density-interference trade-off in multiple-layer LEO satellite IoT constellations: while moderate density enhances coverage through improved satellite availability, excessive density ultimately degrades performance through interference accumulation. This suggests the existence of optimal density configurations that balance these competing effects.
   
  \begin{figure}[h]
   	\centering
    \includegraphics [width=0.4\textwidth] {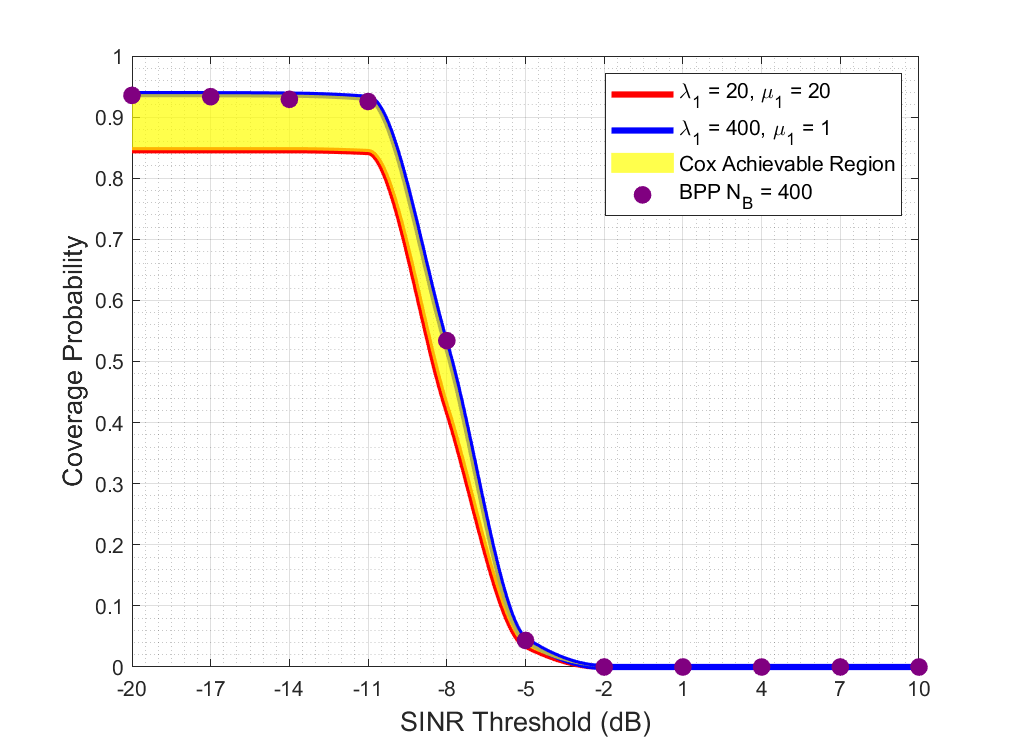}
	\caption {Coverage probability performance for the BPP and CPP models.}
	\label{bpp}
   \end{figure}
   
   To evaluate the effectiveness of the CPP model, we compare it with BPP model under the same total number of LEO satellites. It can be seen in Fig. \ref{bpp} that CPP model can achieve a wider range of adjustment than BPP. Specifically, the BPP models a LEO satellite constellation with only a single parameter, the total number of satellites $N$. However, the CPP model provides two parameters (the number of orbits $\lambda$ and the number of satellites per orbit $\mu$) to describe a constellation of LEO satellites. This stands in sharp contrast to the binomial approach, as a fixed total $N$ (where $N = \lambda\mu$) can be achieved through diverse combinations of orbit counts and satellite densities per orbit. This flexibility is critical for modeling real-world constellations, allowing the CPP model to produce a vast set of potential LEO satellite constellations. Consequently, it enables a far more comprehensive assessment of performance by accounting for a greater representative variety in spatial distribution.
   
   \begin{figure}[h]
   		\centering
   		\includegraphics [width=0.4\textwidth] {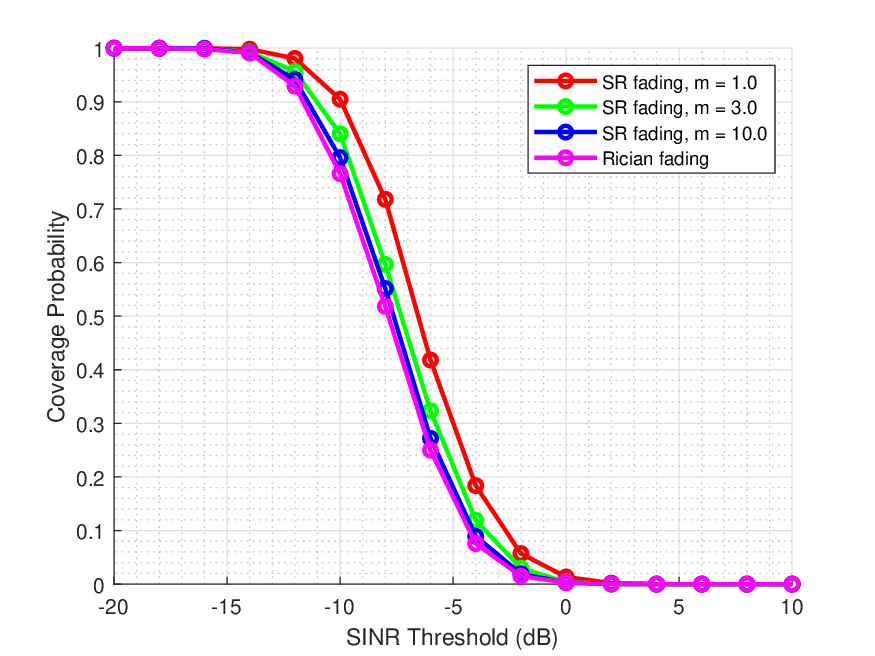}
   		\caption { Coverage probability performance under Rician fading and shadowed-Rician fading with different Nakagami shape parameters.}
   		\label{chancomp}
   \end{figure}
   
   \begin{table}[h]
   	\small
   	\centering
   	\caption{Simulation Parameters of Shadowed-Rician Scenarios}\label{srp1}
   	\begin{tabular}{|c|c|c|c|}
   		\hline
   		Shadowing scenarios & \(b_0\) & \(m\) & \(\Omega\) \\\hline
   		\hline
   		Infrequent light shadowing & 0.158 & 19.4 & 1.29 \\
   		\hline
   		Frequent heavy shadowing  & 0.063 & 0.739 & \(8.97 \times 10^{-4}\) \\
   		\hline
   		Average shadowing         & 0.126 & 10.1 & 0.835 \\
   		\hline
   	\end{tabular}
   \end{table}

\begin{figure}[h]
		\centering
		\includegraphics [width=0.4\textwidth] {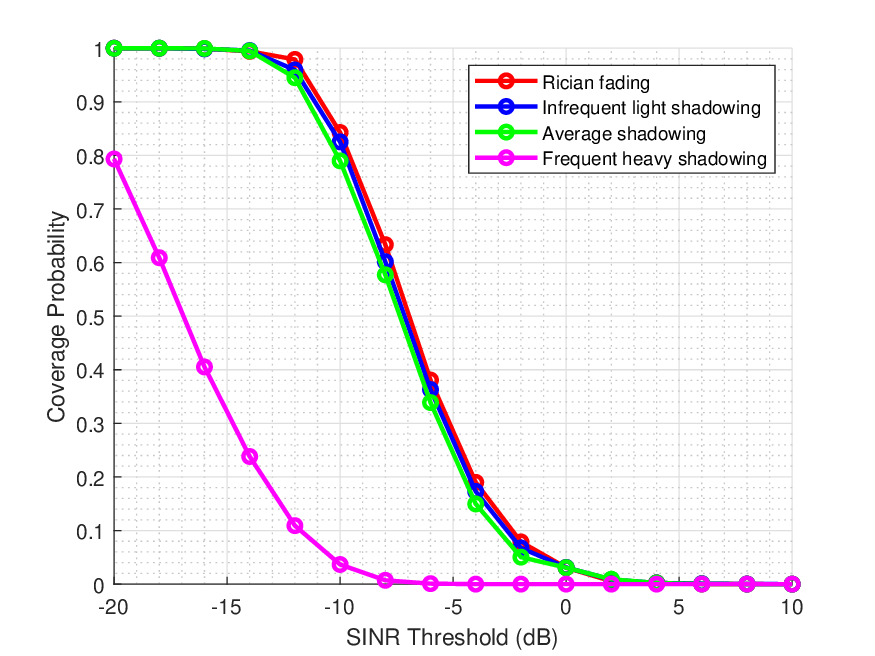}
		\caption {Coverage probability performance under different fading scenarios.}
		\label{chancomp1}
	\end{figure}

      {Finally, to assess the robustness of the proposed framework under diverse propagation conditions, coverage probability is analyzed across both parameterized Shadowed-Rician (SR) fading and established canonical SR scenarios. Fig. \ref{chancomp} compares the coverage probability performance of the proposed framework under Rician fading and SR fading with different severity parameter $m$. A counterintuitive trend is observed: SR fading with smaller $m$ values, which indicates more severe shadowing, yields higher coverage probability. This result stems from the interplay between the maximum received power association criterion and the interference profile in multi-layer LEO satellite constellations. While the serving link is selected based on the most favorable instantaneous channel, severe shadowing simultaneously suppresses the aggregate interference from other LEO satellites, leading to a net improvement in the overall SINR balance. Furthermore, the close alignment between the Rician curve and the SR curve with $m = 10$ confirms the consistency of the analytical framework across fading models. This robustness is also validated under three typical SR scenarios defined in \cite{R1}, namely infrequent light, frequent heavy, and average shadowing, whose parameters are listed in Table \ref{srp1}. As illustrated in Fig. \ref{chancomp1}, the proposed scheme maintains robust performance under infrequent light and average shadowing, matching the performance of a conventional Rician channel. Although frequent heavy shadowing, characterized by significantly diminished LoS power, leads to expected performance degradation, the scheme sustains viable coverage within practical LEO link budgets. Collectively, these analyses demonstrate that the proposed design remains effective across a wide range of realistic fading environments, offering valuable insights for selecting satellite association strategies in varied propagation conditions.}
	
	\section{Discussion}
		{In this section, we firstly discuss three key aspects of the proposed framework to further validate its robustness and practicality: the accuracy of the Rician channel approximation, the generality of the Cox-based model for various constellation designs, and the implications of multi-layer interference and IoT-specific performance.} Finally, we give out the concrete design guidelines for LEO satellite IoT constellations.
	\subsection{Rician Approximation Analysis}
	Firstly, we validate the effectiveness of the proposed finite General Dirichlet series approximation. To establish a rigorous error bound for the proposed approximation, we begin by defining the absolute error function for a given Rician factor $K$ as
	\begin{equation}\label{aef}
		\begin{aligned}
			\epsilon(x;K)&=\left|Q_1\left(\sqrt{2K},\sqrt{\frac{2(1+K)x}{\Omega_s}}\right)-\sum_{n=1}^N\chi_ne^{-\eta_nx}\right|\\
			&=\left|\mathcal{F}'_{|h_{l,m}|^2}(x)-\widetilde{\mathcal{F}}_{|h_{l,m}|^2}'(x)\right|.
		\end{aligned}
	\end{equation}
	Deriving a direct and tight upper bound for $\epsilon(x;K)$ from its explicit form is mathematically challenging. However, the effectiveness of the approximation can be compellingly demonstrated by proving that its error is uniformly small over the domain of interest. Applying the Cauchy-Schwarz inequality to the integrated square error $\mathcal{E}(K)$, yields a worst-case bound on the maximum absolute error: 
	\begin{equation}\label{mae}
		\sup_{x\in[0,x_{\max}]}\epsilon(x;K) \leq \sqrt{\frac{1}{x_{\max}}\cdot\int_0^{x_{\max}}\epsilon(x;K)^{2}dx}\leq\sqrt{\frac{\mathcal{E}(K)}{x_{\max}}},
	\end{equation}
	Given that $\mathcal{E}(K)\approx\hat{\mathcal{E}}(K)$, the numerical error bound $B(K)$ can be obtained as 
	\begin{equation}\label{bk}
		B(K)=\sqrt{\frac{\hat{\mathcal{E}}(K)}{x_{\max}}},
	\end{equation}
	As computed in Table \ref{ChanAPP}, the values of $B(K)$ are consistently very low across all considered $K$-factors. This quantitatively confirms that the maximum pointwise error introduced by our approximation is negligible, thereby validating the high accuracy of the proposed approximation.
	
	Then we focus on the complexity analysis of the proposed approximation method. Specifically, the proposed approximation is analyzed in two phases: offline parameter optimization and online evaluation.
	
	\begin{enumerate}
		\item \textbf{Offline parameter optimization}: The optimization problem Q1 is solved using GA. The complexity depends on the population size $P$, number of generations $G$, and the costs of evaluating $\hat{\mathcal{E}}(K)$ per candidate, which involves computing the Marcum Q-function and the approximation at $i_{\max}=x_{\max}/\delta$. Then, the total complexity is $O(PG
		i_{\max})$. While this is computationally intensive, it is performed once for each $K$, and does not affect online performance. The offline process is feasible on modern workstations.
		
		\item \textbf{Online Evaluation Complexity:} Once the approximation expression is determined, evaluating the approximation for any $x$ requires computing $N$ exponential terms. Thus, the online complexity is $O(N)$, which is extremely efficient compared to direct computation of the Marcum Q-function. For $N=4$, this is negligible in system-level simulations.
	\end{enumerate}
	
	\begin{figure}[h]
		\centering
		\includegraphics [width=0.4\textwidth] {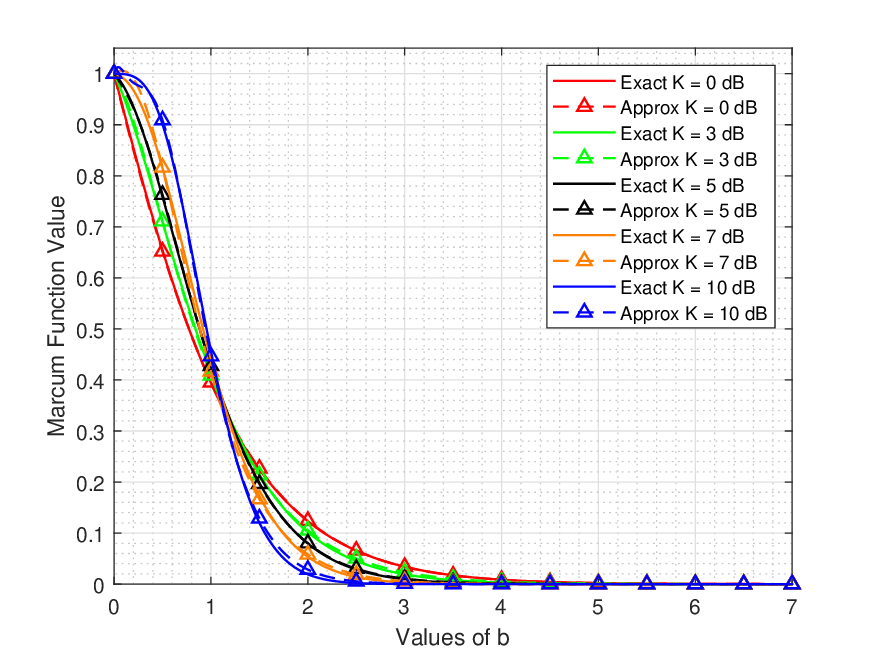}
		\caption { Comparisons with the Marcum Q-function with the proposed approximation method under different Rician factor $K$.}
		\label{Approx}
	\end{figure}

	\begin{table*}
		\small
		\centering
		\caption{Coefficients $\chi_n$ And $\eta_n$ For Different $K$}\label{ChanAPP}
		\begin{tabular}{|c|c|c|c|c|c|c|c|c|c|c|}
			\hline
			\multirow{2}{*}{N}  & \multicolumn{2}{c|}{$K$ = 0 dB} & \multicolumn{2}{c|}{$K$ = 3 dB}  & \multicolumn{2}{c|}{$K$ = 5 dB}& \multicolumn{2}{c|}{$K$ = 7 dB}& \multicolumn{2}{c|}{$K$ = 10 dB}\\  \cline{2-11}
			
			&$\chi_n$ &$\eta_n$ &$\chi_n$ &$\eta_n$  &$\chi_n$ &$\eta_n$ &$\chi_n$ &$\eta_n$  &$\chi_n$ &$\eta_n$ \\ \hline
			$n =1$& -45.90 & 1.86 & 56.32 & 1.83 & 43.13 & 2.90& 28.77 & 3.91& 177.75 & 3.87\\\hline
			$n = 2$& -22.82 & 1.67& 45.12 & 1.81 & 74.79 & 2.50& 95.44 & 3.44 & -338.05 & 4.38\\\hline
			$n = 3$  & 33.44 & 1.92 & -61.05 & 1.82 & 22.34 & 2.92 & 11.77 & 2.53 & 297.00 & 5.40\\\hline
			$n = 4$  & 36.28 & 1.65 & -39.40 & 1.86 & -139.27 & 2.71 & -134.98 & 3.48 & -135.70 & 5.99\\\hline
			$\widehat{\mathcal{E}} (K)$  &\multicolumn{2}{c|}{$8.3954\times 10^{-6}$} & \multicolumn{2}{c|}{$7.4144\times 10^{-5}$} & \multicolumn{2}{c|}{$1.1045\times 10^{-5}$} & \multicolumn{2}{c|}{$1.8885\times 10^{-4}$} & \multicolumn{2}{c|}{$1.5746\times 10^{-4}$} \\\hline
			$B(K)$  &\multicolumn{2}{c|}{$2.8975\times 10^{-3}$} & \multicolumn{2}{c|}{$8.6107\times 10^{-3}$} & \multicolumn{2}{c|}{$3.3234\times 10^{-3}$} & \multicolumn{2}{c|}{$1.3742\times 10^{-2}$} & \multicolumn{2}{c|}{$1.2548\times 10^{-2}$} \\\hline
		\end{tabular}
	\end{table*}

	The accuracy of the proposed approximation is validated across different Rician factors $K$ in Fig.~\ref{Approx} and Table~\ref{ChanAPP}. Fig.~\ref{Approx} demonstrates excellent visual alignment between the approximated and exact Marcum-$Q$ functions, while Table~\ref{ChanAPP} quantitatively confirms the approximation's high accuracy through consistently low error bounds. These results validate the effectiveness of the method in accurately approximating the Marcum-$Q$ function across various channel conditions.

\subsection{Model for Practical Satellite Constellations}

\begin{table}[htbp]
	\scriptsize
	\centering
	\caption{Parameters of Validated LEO Satellite Constellations}
	\label{consval}
	\begin{tabular}{|c|c|c|c|c|}
		\hline
		\textbf{Constellation} & \textbf{Altitude (km)} & \textbf{Planes} & \textbf{Sats/Plane} & \textbf{Inclination (°)}  \\\hline\hline
		
		Starlink (Shell 1) & 530 & 28 & 120 & 43  \\\hline
		Starlink (Shell 2) & 525 & 28 & 120& 53  \\\hline
		Starlink (Shell 3) & 535 & 28 & 120 & 33  \\\hline
		OneWeb (Gen 1) & 1,200 & 12 & 54 & 86.4 \\\hline
		Kuiper (Shell 1) & 630 & 34 & 34 & 51.9 \\\hline
		Kuiper (Shell 2) & 610 & 42 & 36 & 36.0  \\\hline
		Kuiper (Shell 3) & 590 & 28 & 28 & 33.0  \\\hline
		
	\end{tabular}
\end{table}

\begin{figure}[h]
	\centering
	\includegraphics [width=0.4\textwidth] {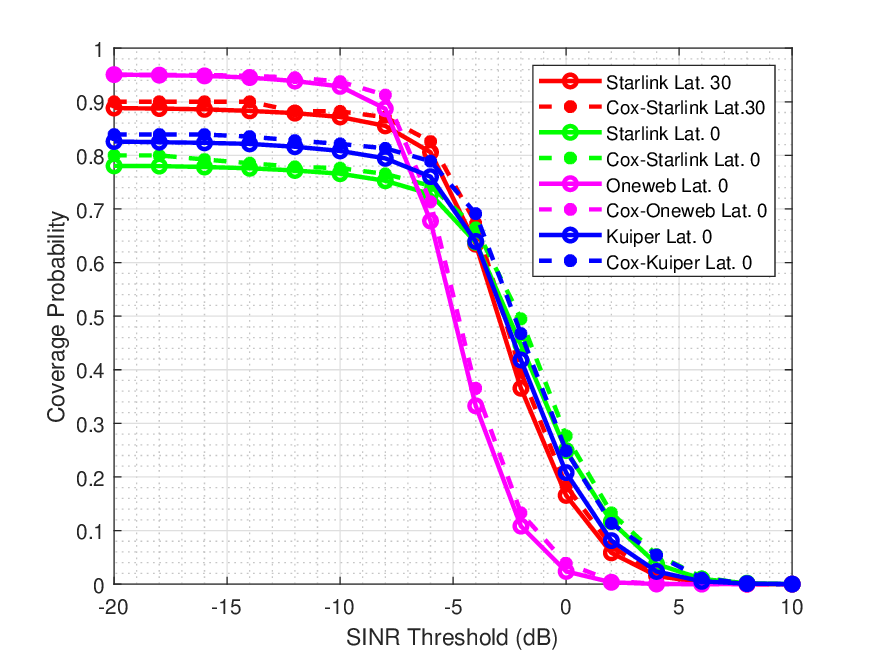}
	\caption {The coverage probability performance of different satellite constellations with the CPP model.}
	\label{sate_cons}
\end{figure}

To establish the robustness and practical utility of the CPP model, we validate the model against a diverse set of constellation architectures. Specifically, we provide a comprehensive performance comparison using the precise orbital parameters of three major LEO constellations with fundamentally different design philosophies: Space X's Starlink, OneWeb, and Amazon's Project Kuiper, as illustrated in Table \ref{consval} \cite{starlink1}. Moreover, the frequency reuse factor is adopted 8 for Starlink and 4 for Kuiper. 

{To represent the orbital distributions and satellites of these constellations by using Cox point processes, we enhance a moment matching method which aligns the number of orbits and the number of satellites of the Cox point process to ensure that the CPP model approximate the local distribution of LEO satellites in those forthcoming constellations.} Specifically, the average number of visible satellites for Starlink, Oneweb, and Kuiper at the latitude of $0^\circ$ is 38, 34 and 37, respectively. Fig. \ref{sate_cons} demonstrates a remarkably close agreement between the CPP model and the high-fidelity simulations across all three heterogeneous architectures. This consistent accuracy, observed despite the vast differences in orbital altitudes, inclination strategies, and shell organizations, provides compelling empirical evidence that the rotation-invariant Cox process effectively captures the essential statistical spatial distribution of satellites from a user's perspective. It robustly confirms that the model's simplifying assumptions do not hinder its ability to deliver accurate, system-level performance estimates for a wide variety of current and planned constellation designs, thereby solidifying its generality and practical value for network analysis and optimization.

Moreover, the visibility of satellites of practical constellations varies based on the user latitudes. However, the CPP model is rotation invariant which means the geometric model is independence from the latitudes. Therefore, we focus on two distinct satellite distributions: one at the equator and another at a latitude of 30 degrees. Specifically, the average number of Starlink satellites visible from latitude $30^\circ$ is 60, while the average number of visible Starlink satellites at the latitude of $0^\circ$ is 38. Based on these empirical mean values, we determine the satellite density $\mu_l$ and orbit density $\lambda_l$ of the CPP model to represent the Starlink constellations by keeping the same average numbers of visible satellites. It is shown in Fig. \ref{sate_cons} that the proposed CPP model match well with the Starlink constellation, which provides guidance for performance analysis of future LEO satellite IoT.

\subsection{Multi-layer and IoT-specific Performance}
Building upon the analytical framework, this subsection delves deeper into the performance implications of multi-layer architectures and their specific relevance to IoT applications. We examine the intricate dynamics of inter-layer interference and explicitly connect the finite blocklength analysis to core IoT requirements based on the short package transmission rate.

\begin{figure}[h]
	\centering
	\includegraphics [width=0.4\textwidth] {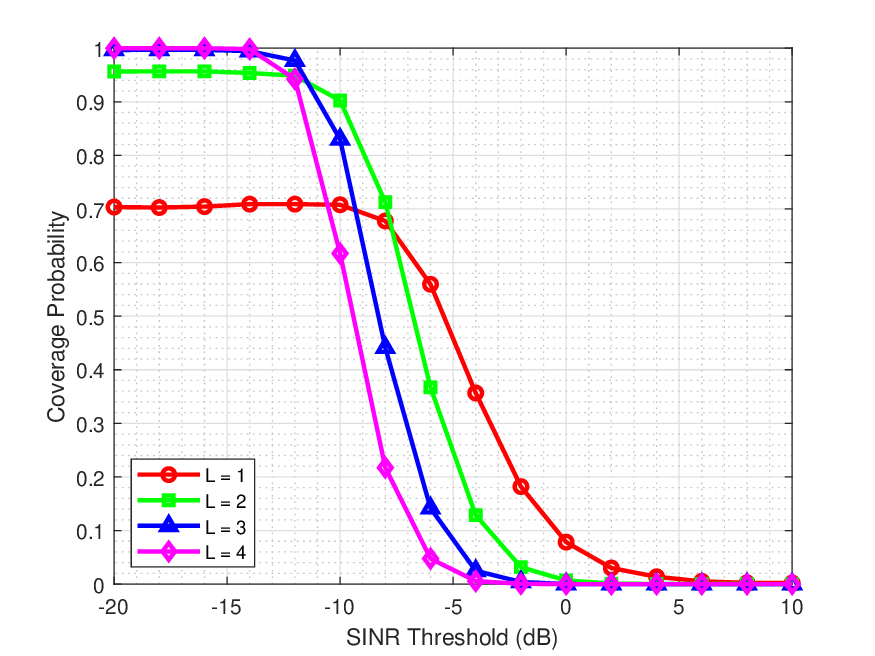}
	\caption { Coverage probability performance under number of layers $L$ for LEO satellite IoT constellation.}
	\label{multilayer}
\end{figure}

As illustrated in Fig. \ref{multilayer}, we evaluate the coverage probability of the LEO satellite constellation under different numbers of layers. It can be observed that increasing the number of layers improves the coverage probability in the low SINR regime, as it enhances the likelihood of a device finding a visible satellite with a strong signal. However, when the satellite coverage becomes sufficient, further increasing the number of layers introduces significant inter-layer interference, which ultimately degrades the coverage performance, particularly in the high SINR region. This reveals a fundamental trade-off between multi-layer diversity and interference accumulation in LEO satellite IoT.

\begin{figure}
	\centering
	\includegraphics [width=0.4\textwidth] {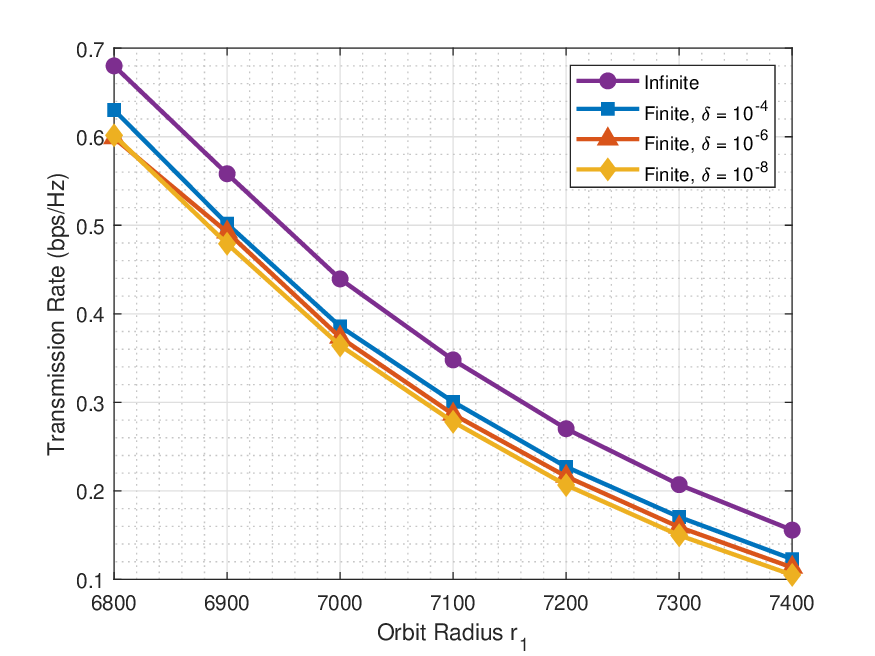}
	\caption {Transmission rate performance under different target error rates versus various orbit radius $r_1$.}
	\label{rate1}
\end{figure}

\begin{figure}
	\centering
	\includegraphics [width=0.4\textwidth] {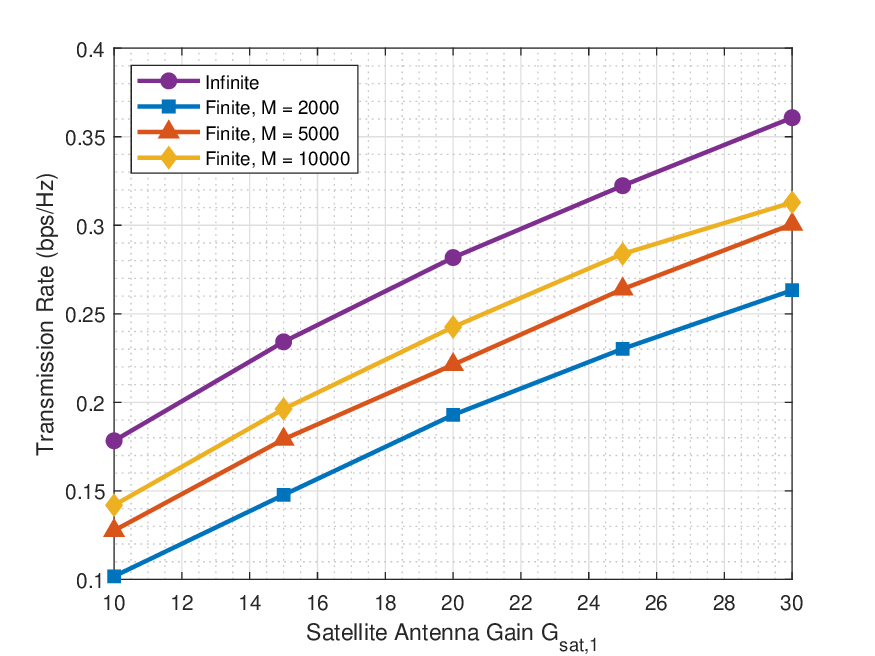}
	\caption {Transmission rate performance under various satellite antenna gain $G_{\mathrm{sat,1}}$ for different block-length $M$.}
	\label{rate2}
\end{figure}

{Then, we focus on the IoT-specific performance in LEO satellite IoT constellation, especially the achievable rate discussed in finite block-length packet transmission.} The analysis of Fig. \ref{rate1} explicitly quantifies the fundamental trade-off between latency and reliability for short-packet IoT communications. The results demonstrate a critical finding for IoT that the achievable rate under practical short-packet constraints is significantly lower than the Shannon capacity, especially at lower orbital altitudes where the potential channel capacity is higher. This ``finite block-length penalty" quantifies the fundamental trade-off between reliability (lower error rate), data rate, and latency (implicit in block-length). This reveals that while lower orbits improve channel conditions, the gains are substantially limited by the short-packet characteristic of IoT traffic, providing crucial guidance for setting realistic performance expectations in LEO satellite IoT constellations.

Moreover, Fig. \ref{rate2} investigates the impact of satellite antenna gain and the block-length on the achievable rate. This analysis directly addresses the practical design space for IoT applications. The results show that increasing the antenna gain provides substantial performance improvements that are consistent across different block-lengths. More importantly, the analysis reveals that for a given antenna gain, increasing the block-length $M$ yields diminishing returns, particularly in the lower gain regime. This provides valuable insight for network operators: while increasing packet duration improves reliability, there is an optimal point beyond which the latency costs outweigh the marginal coding gains. This is particularly relevant for delay-sensitive IoT applications, guiding the joint optimization of physical parameters and transmission strategies.

\subsection{Concrete Design Guidelines}
  The rigorous analytical framework and extensive simulation results yield concrete design guidelines for future multi-layer LEO satellite IoT constellations, directly addressing practical implementation considerations. Constellation designers should target an optimal satellite density that balances connectivity gains against interference effects, as both under-provisioning and over-provisioning can degrade coverage performance. Furthermore, power allocation should be prioritized toward higher-altitude orbital layers, whose transmit power demonstrates greater influence on overall system coverage compared to lower layers. For accurate performance prediction, realistic channel assessment must incorporate Shadowed-Rician fading rather than relying solely on the ideal Rician assumption, particularly in environments susceptible to shadowing where coverage probability is significantly affected.

To effectively support IoT-specific requirements, the finite block-length analysis confirms that lower orbital altitudes combined with substantial satellite antenna gains are essential to establish the necessary SNR conditions for reliable short-packet communications. Additionally, the satellite coverage angle requires careful optimization to balance the competing needs of high signal quality for connected users against broad service availability for potential users across different geographic regions. Collectively, these guidelines provide system architects with a concrete foundation for optimizing future LEO satellite IoT deployments, effectively bridging the gap between our theoretical framework and practical constellation design constraints.	

   \section{Conclusion}
        This paper presented a comprehensive stochastic geometric framework for analyzing multiple-layer LEO satellite IoT constellations, incorporating both Rician fading channels and CPP modeling. In particular, a novel channel approximation method was proposed to effectively simplify complex expressions arising from Rician fading characteristics. Building upon this foundation, we have derived exact theoretical expressions for three fundamental performance metrics: connectivity probability, coverage probability, and transmission rate. Finally, extensive simulation results validated the accuracy and effectiveness of our proposed model. Moreover, the results yield several important design insights that offer valuable guidance for future LEO satellite IoT constellation deployment and optimization.

   \begin{appendices}
   	\section{The Proof of Proposition 1}
        Noticing that the small-scale fading $|h_{l,m}|^2$ follows i.i.d. distribution, thus the Laplace form of $I_k$ can be derived as
    \begin{equation}\label{proof1}
        \begin{aligned}
            \mathcal{L}&_{{I_k}} (s)= \mathcal{L}_{I_k^{\mathrm{Sev}}}(s)\mathcal{L}_{I_k^{\mathrm{Inf}}}(s)\\
           = &  \prod_{Z_i\in \mathcal{Z}_k
           }^{|\varphi_i-\pi/2|\leq \xi_k}\mathbb{E}\Bigg[\prod_{m'\in \psi_{k,i}} \mathbb{E}_{|h|^2}\Big[e^{-sg_tp_k G_{\mathrm{sat},k}|h|^2 d_{k,m'}^{-2}} \Big]\Bigg] \cdot\\
           &\prod_{Z_i\in \mathcal{Z}_k
           }^{\xi_k\leq|\varphi_i-\pi/2|\leq \overline{\varphi}_{\beta, k}}\mathbb{E}\Bigg[\prod_{m'\in \psi_{k,i}} \mathbb{E}_{|h|^2}\Big[e^{-sg_tp_k G_{\mathrm{sat},k}|h|^2 d_{k,m'}^{-2}} \Big]\Bigg]\cdot\\
          & \prod_{Z_i\in\mathcal{Z}_k}^{|\varphi_i-\pi/2|\leq \overline{\varphi}_{\beta, k}}\mathbb{E}\Bigg[\prod_{m'\in \overline{\psi}_{k,i}} \mathbb{E}_{|h|^2}\Big[e^{-sg_tp_k |h|^2 d_{k,m'}^{-2}} \Big]\Bigg] \cdot \\
          & \prod_{Z_i\in\mathcal{Z}_k}^{\overline{\varphi}_{\beta, k}\leq|\varphi_i-\pi/2|\leq \overline{\varphi}_{ k}}\mathbb{E}\Bigg[\prod_{m'\in \overline{\psi}_{k,i}} \mathbb{E}_{|h|^2}\Big[e^{-sg_tp_k |h|^2 d_{k,m'}^{-2}} \Big] \Bigg],
        \end{aligned}
    \end{equation}
    where $d_{k,m'}$ is the distance between the LEO satellite $m'$ of layer $k$ and the typical IoT device. According to \cite{cox1}, $d_{k,m'}$ can be expressed as
    \begin{equation}\label{dkm}
        d_{k,m'} = \sqrt{r_k^2-2 r_{k} r_e\cos \left(\omega_{m'}\right) \sin \left(\varphi_{i}\right)+r_{e}^{2}}=f_{k,\varphi_i}(\omega_{m'}).
    \end{equation}
      $\xi_k$, $\overline{\varphi}_{\beta,k}$ and $\overline{\varphi}_{k}$ are given by 
    \begin{equation}\label{xi_k}
        \xi_k = \arccos\left(\frac{r_e^2+r_k^2-d^2}{2r_er_k}\right),
    \end{equation}
    \begin{equation}\label{pbk}
        \overline{\varphi}_{\beta,k}=\arccos\left(\frac{r_k}{r_e}\sin^2\beta_k-\sqrt{1-\frac{r_l^2}{r_e^2}\sin^2\beta_k}\cos\beta_k\right),
    \end{equation}
    \begin{equation}\label{pk}
        \overline{\varphi}_{k}=\arccos\frac{r_e}{r_k}.
    \end{equation}
    Then, by applying PGFL of a PPP of intensity $\mu_k$ \cite{prf1}, \cite{prf2}, the Laplace form $\mathcal{L}_{|h|^2}(s)=\mathbb{E}_{|h|^2}[e^{-s|h|^2}]$ and (\ref{dkm}), (\ref{proof1}) can be further rewritten as
    \begin{equation}\label{proof2}
        \begin{aligned}
        \mathcal{L}&_{{I_k}} (s)=   \prod_{Z_i\in \mathcal{Z}_k
           }^{|\varphi_i-\pi/2|\leq \xi_k}e^{\frac{\mu_k}{\pi}
            \int_{w_{l,\varphi_i,1}}^{w_{k,\varphi_i,2}}
            1-\mathcal{L}_{|h|^2}\left(\frac{sg_tp_kG_{\mathrm{sat},k}}{f_{l,\varphi_i}^2(w)}\right)d w }\cdot\\
            & \prod_{Z_i\in \mathcal{Z}_k
           }^{\xi_k\leq|\varphi_i-\pi/2|\leq \overline{\varphi}_{\beta, k}}e^{-\frac{\mu_k}{\pi}
            \int_{0}^{w_{k,\varphi_i,2}}1-\mathcal{L}_{|h|^2}\left(\frac{sg_tp_kG_{\mathrm{sat},k}}{f_{l,\varphi_i}^2(w)}\right)d w}\cdot\\  &\prod_{Z_i\in\mathcal{Z}_k}^{ |\varphi_i-\pi/2|\leq \overline{\varphi}_{\beta, k}}e^{-\frac{\mu_k}{\pi}\int_{w_{k,\varphi_i,2}}^{w_{k,\varphi_i,3}}1-\mathcal{L}_{|h|^2}\left(\frac{sg_tp_1}{f_{k,\phi_i}^2(w)}\right)d w} \cdot\\
           & \prod_{Z_i\in \mathcal{Z}_k
           }^{\overline{\varphi}_{\beta, k}\leq|\varphi_i-\pi/2|\leq \overline{\varphi}_{k}}e^{-\frac{\mu_k}{\pi}
            \int_{0}^{w_{k,\varphi_i,3}}1-\mathcal{L}_{|h|^2}\left(\frac{sg_tp_k}{f_{l,\varphi_i}^2(w)}\right)d w} .   
        \end{aligned}
    \end{equation}
    where we have 
    \begin{equation}\label{w1}
        w_{k,\varphi_i,1} = \arcsin(\sqrt{1-\cos^2\xi_k\csc^2\varphi_i}),
    \end{equation}

    \begin{equation}\label{w2}
        w_{k,\varphi_i,2} = \arcsin(\sqrt{1-\cos^2\overline{\varphi}_{\beta,k}\csc^2\varphi_i}),
    \end{equation}
    
    \begin{equation}\label{w3}
        w_{k,\varphi_i,3} = \arcsin(\sqrt{1-\cos^2\overline{\varphi}_{k}\csc^2\varphi_i}).
    \end{equation}
    Finally, by rearranging the terms in (\ref{proof2}), we can obtain the results in (\ref{lpm}).
    
    The proof is completed.

\end{appendices}

\end{document}